\documentclass[11pt,draftclsnofoot,onecolumn]{IEEEtran}
\ifCLASSINFOpdf

\else

\fi

\usepackage{amssymb}  
\usepackage{graphics} 
\usepackage{epsfig} 
\usepackage{mathptmx} 
\usepackage{times} 
\usepackage{amsmath} 
\usepackage{booktabs}
\usepackage{threeparttable}
\usepackage{floatrow}
\usepackage{multirow}
\usepackage{subfigure,url}
\usepackage{hyperref}
\usepackage{makecell,xcolor,bm}
\newfloatcommand{capbtabbox}{table}[][\FBwidth]
\DeclareMathAlphabet{\mathcal}{OMS}{cmsy}{b}{n}
\DeclareMathAlphabet{\mathcal}{OMS}{cmsy}{m}{n}

\newtheorem{theorem}{\indent Theorem}
\newtheorem{lemma}{\indent Lemma}
\newtheorem{corollary}{\indent Corollary}
\newtheorem{proposition}{\indent Proposition}
\newtheorem{definition}{\indent Definition}

\newtheorem{remark}{\indent Remark}
\newtheorem{problem}{\indent Problem}

\begin{document}
\title{Simultaneous estimations of quantum state and detector through multiple quantum processes}
\author{Shuixin Xiao, Weichao Liang, Yuanlong Wang, Daoyi Dong, Ian R. Petersen and Valery Ugrinovskii$^*$

\thanks{This research was supported by the Australian Research Council (DP200102945, DP210101938, FT220100656) and the National Natural Science Foundation of China (12288201). Shuixin Xiao would like to gratefully acknowledge the support from the IEEE Control Systems Society Graduate Collaboration Fellowship.}
\thanks{Shuixin Xiao is with School of Engineering and Technology, University of New South Wales, Canberra ACT 2600, Australia, and School of Engineering, Australian National University, Canberra, ACT 2601, Australia (e-mail: shuixin.xiao@unsw.edu.au). }
\thanks{Weichao Liang is with School of Engineering and Technology, University of New South Wales, Canberra ACT 2600, Australia (email: weichao.liang@unsw.edu.au).}
\thanks{Yuanlong Wang is with Key Laboratory of Systems and Control, Academy of Mathematics and Systems Science, Chinese Academy of Sciences, Beijing 100190, China (e-mail: wangyuanlong@amss.ac.cn).}
\thanks{Daoyi Dong and Ian R. Petersen are with CIICADA Lab, School of Engineering, Australian National University, ACT 2601, Australia (e-mail: daoyi.dong@anu.edu.au, i.r.petersen@gmail.com).}
\thanks{$^*$Valery Ugrinovskii was with School of Engineering and Technology, University of New South Wales, Canberra ACT 2600, Australia.  His sudden passing left an irreplaceable void in our field. We thank him for his invaluable contributions.}
	}

\maketitle

\begin{abstract}
	The estimation of all the
	parameters in an unknown quantum state or measurement device, commonly known as quantum state tomography (QST) and quantum detector tomography (QDT), is crucial for comprehensively characterizing and controlling quantum systems. In this paper, we introduce a framework,  in two different bases, that utilizes multiple quantum processes to simultaneously identify a quantum state and a detector.
	We develop a closed-form algorithm for this purpose and prove that the mean squared error (MSE) scales as $O(1/N) $ for both QST and QDT, where $N $ denotes the total number of state copies. This scaling aligns with established patterns observed in previous works that addressed QST and QDT as independent tasks.
	Furthermore, we formulate the problem as a sum of squares (SOS) optimization problem with semialgebraic constraints, where the physical constraints of the state and detector are characterized by polynomial equalities and inequalities. The effectiveness of our proposed methods is validated through numerical examples.

\end{abstract}

\section{Introduction}
Quantum system identification \cite{PhysRevLett.108.080502,7130587} and quantum tomography \cite{qci,dong2022quantum} are essential endeavors for obtaining comprehensive models of quantum systems. This pursuit is crucial for the thorough exploration and management of quantum systems \cite{ wiseman2009quantum,Dong2023}, facilitating advancements in various quantum science applications such as quantum computation \cite{qci}, quantum sensing \cite{RevModPhys.89.035002}, and quantum control \cite{Zhang2017,Liang2024}.
In quantum tomography, the primary focus often revolves around two key challenges: quantum state tomography (QST) and quantum detector tomography (QDT). These tasks involve the estimation of all the parameters associated with an unknown quantum state and detector.

For QST, various algorithms have been developed, including Maximum Likelihood Estimation (MLE) \cite{qstmle,effqst} and Linear Regression Estimation (LRE) \cite{Qi2013,xiaorank}. Innovative approaches have emerged for low-rank quantum states, such as the application of compressed sensing \cite{Gross2010,Flammia2012}. Additionally, regularization techniques have been introduced to enhance the QST accuracy \cite{MU2020108837}. To predict many properties of quantum states with few measurements, shadow tomography has been proposed \cite{Aaronson2018,Huang2020}.
For QDT, the pioneering approach involved MLE \cite{PhysRevA.64.024102}. Subsequent methodologies include linear regression \cite{Grandi_2017}, function fitting \cite{Renema2012}, and convex optimization \cite{Feito_2009, Lundeen2009, Zhang_2012}. A recent advance in \cite{wang2019twostage} introduced a two-stage estimation approach characterized by analytical computational complexity and an upper bound for the mean squared error (MSE). Building upon this approach, optimization of probe states was introduced \cite{xiao2021optimal}, and regularization techniques were studied \cite{Xiao2023}. Self-calibration and direct characterization of a detector using weak values were also investigated in \cite{PhysRevLett.124.040402,PhysRevLett.127.180401}.

Existing works primarily focus on performing QST and QDT as separate tasks. Typically, precise measurement devices are assumed for QST, while QDT relies on known quantum states and measurement results to estimate the unknown detector. However, real-world scenarios inevitably involve State Preparation and Measurement (SPAM) errors, which can impair the accuracy of quantum tomography. 
In this paper, we propose a framework to simultaneously identify an unknown quantum state and an unknown detector using multiple quantum processes.
 Our approach does not depend on prior assumptions regarding the unknown state or detector, making it independent of SPAM errors.
The approach involves inputting the same unknown quantum state into multiple known quantum processes and applying the same unknown detector  on the output states, as illustrated in Fig.~\ref{f1}. Leveraging the obtained measurement results enables the simultaneous identification of the input state and detector. Previous work is restricted to unitary processes \cite{PhysRevA.98.042318} or single-qubit systems \cite{PhysRevA.104.012416}. However, our framework can be implemented for arbitrary-dimension quantum systems and for generalized-unital processes, of which unitary processes constitute a special case, or for arbitrary processes.

When the quantum processes are generalized-unital, we formulate the task of simultaneously identifying an unknown quantum state and an unknown detector as an optimization problem based on the orthonormal operator basis. For arbitrary quantum processes, we formulate the task as an optimization problem using the natural basis. We then present closed-form solutions for these two optimization problems and subsequently analyze the MSE scalings of our algorithm.
 The MSE scalings of QST and QDT are both $O(1/N)$ in the informationally complete scenario, where $N$ denotes the total number of state copies. These scalings are consistent with previous works that treated the tasks of QST and QDT separately, as demonstrated in \cite{Qi2013,wang2019twostage}.
We further formulate the problem as a sum of squares (SOS) optimization problem with semialgebraic constraints and utilize SOSTOOLS \cite{sostools} to solve it. SOSTOOLS provides a lower bound on the cost function and may not always yield the values of the optimization variables. However, if SOSTOOLS provides the values of the optimization variables achieving the lower bound, this lower bound represents the minimum value of the cost function. Our study is the inaugural endeavor to implement SOS optimization techniques in quantum tomography, marking a significant advancement in the field.
In addition, we explore several illustrative scenarios, including using closed quantum systems driven by multiple Hamiltonians, applying mixed-unitary quantum processes, and identifying pure input states.
Finally, we present numerical results to validate the effectiveness of the closed-form solution and SOS optimization. We find that the closed-form solution is fast but with low accuracy, while SOS optimization achieves higher accuracy at the cost of longer computation time.

The main contributions
of this paper are summarized as follows.
\begin{enumerate}
	\item[(i)]  We propose a framework for simultaneously identifying a quantum state and detector independent of SPAM errors, formulating the task into two versions of optimization problems in two different bases. This framework allows for non-unitary processes to be employed in arbitrary-dimension systems.
	\item[(ii)] We provide a closed-form solution to the optimization problems with proved MSE scalings $O(1/N)$. Additionally, we formulate the problem as an SOS optimization problem with semialgebraic constraints. Our work represents the first implementation of SOS optimization techniques in quantum tomography.
	\item[(iii)] We explore several illustrative scenarios, including multiple Hamiltonians in closed systems, mixed-unitary processes, and   pure input states.
	\item[(iv)] Numerical examples are conducted to validate the theoretical results on the MSE scalings and demonstrate the effectiveness of our method.
\end{enumerate}

The organization of this paper is as follows. Section
\ref{sec2} presents a problem formulation to implement QST and QDT simultaneously. Section \ref{closed} proposes a closed-form algorithm, and proves the corresponding  MSE scalings. SOS optimization is discussed in Section \ref{sosop}. 
Several illustrative examples and
numerical simulation results are presented in Section \ref{secexample}  and Section \ref{sec6}, respectively.  Section \ref{sec7} concludes this paper. 

Notation:  The $ i $-th row and $ j $-th column of a matrix $ X $ is $(X)_{ij} $. The elements from the $m$-th to the $n$-th position ($m\leq n$) in a vector $x$ are denoted as $x_{m:n}$. The transpose of $X$ is $X^T$. The conjugate $ (*) $ and transpose of $X$ is $X^\dagger$. The rank of a matrix $X$ is $\operatorname{rank}(X)$. 
The sets of real and complex numbers are $\mathbb{R}$ and $\mathbb{C}$, respectively. The sets of  $d$-dimension complex vectors and  $d\times d$ complex matrices are $\mathbb{C}^d$ and $\mathbb{C}^{d\times d}$, respectively. The identity matrix is $ I $. The zero vector in $\mathbb{R}^{m}$ is $0_m$.
$\rm i=\sqrt{-1}$. 
The trace of $X$ is $\text{Tr}(X)$. The Frobenius norm of a matrix $X$ is denoted as $||X||$ and the 2-norm of a vector $ x $ is $||x||$.   The estimate of $X$ is $ \hat X $. The inner product of two matrices $X$ and $Y$ is defined as $\langle X, Y\rangle\triangleq\text{Tr}(X^\dagger Y)$. The inner product of two vectors $x$ and $y$ is defined as $\langle x,
y\rangle\triangleq x^\dagger y$.
 The tensor product of $A$ and $B$ is denoted $A\otimes B$. The Kronecker delta function is $\delta$. Denote standard basis as $ \{|i\rangle\}_{i=1}^n $  such that $ \langle i| j\rangle=\delta_{ij}$. The $\operatorname{diag}(a)$ denotes a diagonal matrix with the $i$-th diagonal element being the $i$-th element of the vector $a$. Pauli matrices are $ \sigma_{x} $, $ \sigma_{y} $ and $\sigma_{z}$. The smallest integer not smaller than $x \in \mathbb R$ is given by $\lceil x \rceil$.


\section{Problem formulation}\label{sec2}
In this section, we first discuss some preliminary knowledge and define generalized-unital quantum processes. Then, we propose the first version of the problem formulation for generalized-unital processes based on the orthonormal basis and the second version for arbitrary quantum processes using the natural basis. Finally, we present comments on these problem formulations and define informationally complete/incomplete scenarios, highlighting that several common processes are always informationally incomplete.
\subsection{Preliminary knowledge}
\begin{figure}
	\centering
	\includegraphics[width=3.6in]{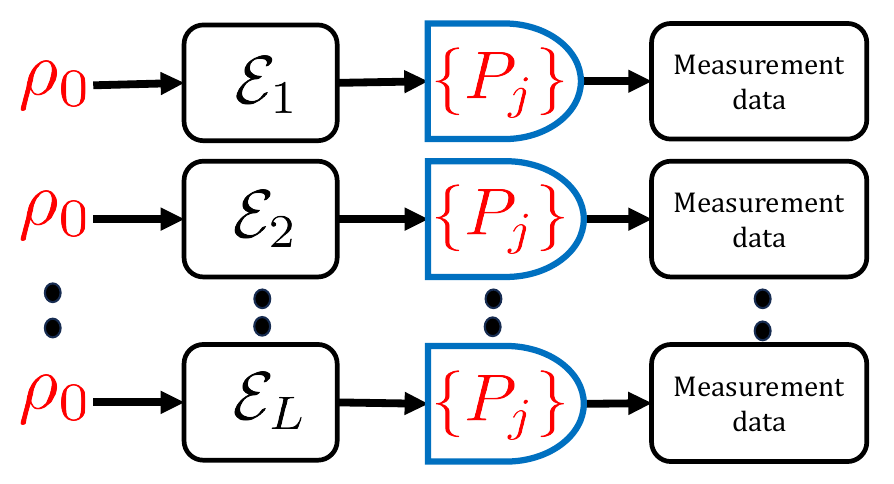}
	\centering{\caption{Schematic diagram to identify the quantum state and detector simultaneously using multiple known quantum processes. We input the same unknown quantum state $\rho_0$ into multiple known quantum processes $\{\mathcal{E}_a\}_{a=1}^{L}$ and apply the same unknown detector $\{P_j\}_{j=1}^{M}$ to obtain measurement data.}\label{f1}}
\end{figure}
For a matrix $A_{m\times n}$, we introduce vectorization function:
\begin{equation}\label{vec}
	\begin{aligned}
		\operatorname{vec}(A_{m\times n})\triangleq&[(A)_{11},(A)_{21},\cdots,(A)_{m1},(A)_{12},\cdots,(A)_{m2},\\
		&\cdots,(A)_{1n},\cdots,(A)_{mn}]^T.
	\end{aligned}
\end{equation}
Similarly, $\text{vec}^{-1}(\cdot)$ maps a $d^2\times 1$ vector into a $d\times d$ square matrix.
The common properties of $ \text{vec}(\cdot) $ are listed as follows \cite{watrous2018theory}:
\begin{equation}\label{property1}
	\langle X, Y\rangle=\langle \text{vec}(X), \text{vec}(Y)\rangle,
\end{equation}
\begin{equation}\label{property2}
	\operatorname{vec}(ABC)=(C^T\otimes A)\operatorname{vec}(B).
\end{equation}

A quantum state of a $d$-dimensional system can be characterized by a density operator $\rho$  belonging to the space
\begin{equation}
	\mathcal{P}=\left\{\rho \in \mathbb{C}^{d\times d}:  \rho=\rho^{\dagger}, \rho\geq 0,\operatorname{Tr}(\rho)=1 \right\}.
\end{equation}
We use a unit complex vector  $|\psi\rangle$ to represent a pure state and its corresponding density operator is $\rho=|\psi\rangle\langle\psi|$.

In quantum physics, measurement plays a fundamental role, and the device responsible for measurement is known as a detector. This detector can be characterized by a set of measurement operators denoted as  $\{P_j\}_{j=1}^{M}$. These operators collectively form a Positive-Operator-Valued Measure (POVM), where each POVM element $P_j \in \mathbb{C}^{d \times d}$ adheres to the conditions $ P_j = P_j^{\dagger} $ and $ P_j \geq 0 $. Additionally, they satisfy the completeness constraint $\sum_{j=1}^M P_j = I$. Therefore, the set of all such $\{P_j\}_{j=1}^{M}$ belongs to the set $\mathcal{R}$ defined as:
\begin{equation}
	\mathcal{R}=\Big\{\{P_j\}_{j=1}^{M}: P_j \in \mathbb{C}^{d\times d}, P_j=P_j^{\dagger}, P_j\geq 0, \sum_{j=1}^M P_j = I \Big\}.
\end{equation}
A widely used measurement type is the projective measurement, where each $P_j=|\phi_j\rangle\langle\phi_j|$ and $|\phi_j\rangle $ is a unit complex vector. Some commonly used projective measurements in quantum tomography include the Symmetric Informationally Complete POVM \cite{sic}, Mutually Unbiased Bases measurements \cite{mubreview}, and Cube bases \cite{PhysRevA.78.052122}.

When a POVM element $P_j$ is applied to a quantum state $\rho$, the probability of obtaining the corresponding result is governed by Born's rule \cite{qci}:
\begin{equation}
	p_{j} = \operatorname{Tr}\left(P_{j} \rho\right).
\end{equation}
From the completeness constraint, we have $\sum_{j=1}^{M} p_j=1$. In practical experiments, suppose that $N$ identical copies of $\rho$ are prepared, and the $j$-th result occurs $N_j$ times. Then $\hat p_j=N_j/N$ serves as the experimental estimate of the true value $p_j$, with the associated measurement error denoted as $ e_{j} = \hat p_{j} - p_{j} $ \cite{wang2019twostage}. According to the central limit theorem, the distribution of $ e_{j} $ converges to a normal distribution with mean zero and variance $(p_{j} - p_{j}^{2})/N$ \cite{Qi2013, MU2020108837}.

For a $ d $-dimensional quantum system,
suppose there are $L$ different quantum processes $\{\mathcal{E}_a\}_{a=1}^{L}$ which are completely-positive (CP) linear maps. For all the processes, the initial states are all $\rho_0$. Then we implement the same detector $\{P_j\}_{j=1}^{M}$ on the output states and obtain the measurement data. Using these measurement data, we aim to concurrently identify the same initial quantum state $\rho_0$ and detector $\{P_j\}_{j=1}^{M}$ with multiple quantum processes as Fig. \ref{f1}. 
Using Kraus operators, for the $a$-th process $\mathcal{E}_{a}$, the output state is
\begin{equation}\label{rhoout}
	\rho_a=\mathcal{E}_{a}(\rho_0)=\sum_{i=1}^{d^2} A_i^a \rho_0\left(A_i^a\right)^{\dagger},
\end{equation}
where $\{A_i^a\}_{i=1}^{d^2} \in \mathbb{C}^{d\times d}$ are the Kraus operators of the $a$-th quantum process. 
These Kraus operators satisfy
\begin{equation}\label{aleq}
\sum_{i=1}^{d^2} ( A_{i}^a)^{\dagger}  A_{i}^a\leq I.
\end{equation}
When the equality in \eqref{aleq} holds, the map $ \mathcal{E}  $ is trace-preserving (TP). Otherwise, it is non-trace-preserving (non-TP). In our framework, the quantum process can be TP or non-TP, and thus $\operatorname{Tr}(\rho_{a})$ may be smaller than one. A quantum process $\mathcal{E}_a$ is called unital if $\mathcal{E}_a(I)=I$  \cite{Mendl2009}, i.e., 
\begin{equation}
\sum_{i=1}^{d^2} A_{i}^a  (A_{i}^a)^{\dagger}= I.
\end{equation}
Here we extend the property of unital to \emph{generalized-unital} which is defined as follows.
\begin{definition}\label{defuni}
	A quantum process $\mathcal{E}_a$ is called \emph{generalized-unital} if $\mathcal{E}_a(I)=\alpha I$ where $0<\alpha \leq 1$ is a constant.
\end{definition}

Thus, any unital process also belongs to generalized-unital processes.
If $\mathcal{E}_a$ is generalized-unital,  we have 
 \begin{equation}
 	\sum_{i=1}^{d^2} A_{i}^a  (A_{i}^a)^{\dagger}= \alpha I.
 \end{equation}
 In Section \ref{secexample}, we will discuss unitary processes and mixed-unitary processes which are both generalized-unital processes.

\subsection{First version of problem formulation}
In this subsection, we focus on generalized-unital processes.
Let $\{\Omega_{j}\}_{j=0}^{d^{2}-1}$ be a complete basis set of orthonormal operators with dimension $d$, satisfying $\operatorname{Tr}(\Omega_{i}^{\dagger} \Omega_{j}) = \delta_{i j}$. Each operator $\Omega_{j}$ is Hermitian, and $\operatorname{Tr}(\Omega_{j}) = 0$ for all $j$ except $\Omega_{0} = I / \sqrt{d}$. Consequently, $\left\{\mathrm{i} \Omega_j\right\}_{j=1}^{d^{2}-1}$ forms an orthonormal basis for the Lie algebra $\mathfrak{su}(d)$.
Let the initial input state be $ \rho_0 $ which can be  parameterized as
\begin{eqnarray}\label{rho}
	\rho_0 =\frac{1}{\sqrt{d}}\Omega_{0}+\sum_{k=1}^{d^2-1} x_{0,k}\Omega_k,
\end{eqnarray}
and denote
\begin{equation}\label{x0}
 x_0\triangleq[x_{0,1}, \cdots, x_{0,d^2-1}]^{T}.
\end{equation}
We also denote the inverse map from $x_0$ to $\rho_0$ as $h(\cdot):  \mathbb{R}^{d^2-1}\rightarrow \mathbb{C}^{d\times d}$.
Let $x_{a,j}\triangleq \operatorname{Tr} \left[\Omega_j \rho_{a}\right]$. Therefore, $x_{a,0}=\operatorname{Tr}(\rho_a)/\sqrt{d} \leq \frac{1}{\sqrt{d}}$ because we consider both TP and non-TP processes. A real vector $ x_{a}\triangleq[x_{a,1}, \cdots, x_{a,(d^2-1)}]^{T} $  representing a quantum state is usually referred to as the coherence vector \cite{zhang1,alicki} for TP processes.

For the $j$-th POVM element $P_j$, it can be  parameterized as
\begin{equation}
	P_j =C_{j,0} \Omega_0+\sum_{k=1}^{d^2-1} C_{j,k} \Omega_k.
\end{equation} 
We denote
\begin{equation}\label{cj}
C_{j}\triangleq[C_{j,1}, \cdots, C_{j,d^2-1}]^{T}.
\end{equation}
as the main part of the parameterization of $P_j$ under $\{\Omega_j\}_{j=0}^{d^2-1}$.
Define
\begin{equation}\label{change}
	U\triangleq\left[\operatorname{vec}(\Omega_{0}), \cdots, \operatorname{vec}(\Omega_{d^2-1})\right]^{\dagger},
\end{equation}
which is unitary and is the change of basis matrix between $\{\Omega_{j}\}_{j=0}^{d^2-1}$ and natural basis $\{|l\rangle\langle k|\}_{1\leq l,k\leq d}$ \cite{8022944}. Using \eqref{vec} and \eqref{change}, we have 
\begin{equation}\label{real}
	\begin{aligned}
			&\left[\begin{array}{c}
		x_{a,0}\\
		x_a
	\end{array}\right]=U\operatorname{vec}\left(\rho_a\right), 	\;
	\left[\begin{array}{c}
		1/\sqrt{d} \\
		x_0
	\end{array}\right]=U\operatorname{vec}\left(\rho_0\right), \\ &\left[\begin{array}{c}
	C_{j,0}\\
	C_{j}
	\end{array}\right]=U\operatorname{vec}\left(P_j\right).
		\end{aligned}
\end{equation}
We then present the following proposition and the proof is given in Appendix \ref{ab}.
\begin{proposition}\label{preal}
	For each matrix $A\in \mathbb{C}^{d\times d}$, $U(A^{*}\otimes A) U^{\dagger}$ is a real matrix.
\end{proposition}

Using \eqref{property2} and \eqref{rhoout}, we have
\begin{equation}\label{eq8}
	\operatorname{vec}\left(\rho_a\right)=\Big(\sum_{i=1}^{d^2}\left(A_i^a\right)^* \otimes A_i^a\Big) \operatorname{vec}\left(\rho_0\right).
\end{equation}
Then
using \eqref{real} and \eqref{eq8}, we have
\begin{equation}\label{xa}
	\begin{aligned}
\left[\begin{array}{c}
x_{a,0} \\
	x_a
\end{array}\right]&=	U\operatorname{vec}\left(\rho_a\right)\\
&=U\Big(\sum_{i=1}^{d^2}\left(A_i^a\right)^* \otimes A_i^a\Big) U^{\dagger} U\operatorname{vec}\left(\rho_0\right)\\
&=U\Big(\sum_{i=1}^{d^2}\left(A_i^a\right)^* \otimes A_i^a\Big) U^{\dagger}
\left[\begin{array}{c}
	1/\sqrt{d} \\
	x_0
\end{array}\right].
	\end{aligned}
\end{equation}    
Using Proposition \ref{preal}, we know $U\left(\sum_{i=1}^{d^2}\left(A_i^a\right)^* \otimes A_i^a\right) U^{\dagger}$ is also a real matrix.  
We partition it as
\begin{equation}\label{xa1}
	U\Big(\sum_{i=1}^{d^2}(A_i^a)^* \otimes A_i^a\Big) U^{\dagger}=	\left[\begin{array}{cc}
		r_a & t_a^{T} \\
		h_a & E_a
	\end{array}\right]
\end{equation}
where $E_a \in \mathbb{R}^{(d^2-1) \times (d^2-1)}$  and $h_a, t_a \in \mathbb{R}^{d^2-1}$.

In this complete basis $\{\Omega_{j}\}_{j=0}^{d^{2}-1}$, we are interested in 
 one special case where $h_a=0_{d^2-1}$ for $1\leq a \leq L$, which in fact covers many common processes and leads to Problem \ref{problem1}.
We propose the following result to characterize this scenario and the proof is presented in Appendix \ref{ac}.
\begin{theorem}\label{p3}
	A process $\mathcal{E}_a$ is generalized-unital if and only if $h_a=0_{d^2-1}$.
\end{theorem}

Using \eqref{xa}, \eqref{xa1} and Theorem \ref{p3}, we have
\begin{equation}\label{eq25}
	x_a=E_a x_0.
\end{equation} 
Therefore,    the ideal measurement data of the $j$-th POVM element $P_j$ on $\rho_a$ is                         
\begin{equation}\label{yaj}
	\begin{aligned}
		y_{aj}&=\operatorname{Tr}\left(P_j \rho_a\right)=\operatorname{vec}\left(P_{j}\right)^{\dagger} \operatorname{vec}\left(\rho_a\right)\\
		&=\operatorname{vec}\left(P_{j}\right)^{\dagger} U^{\dagger} U \operatorname{vec}\left(\rho_a\right)\\
		&=[C_{j,0},  C_j^{T}]\left[\begin{array}{c}
			x_{a,0} \\
			x_a
		\end{array}\right]\\
		&=C_{j,0}  x_{a,0}+C_j^{T} E_a  x_0.
	\end{aligned}
\end{equation}
Using \eqref{property2} and \eqref{yaj}, we have
\begin{equation}
	{Y}_{aj}\triangleq y_{aj}-C_{j,0}  x_{a,0} =\operatorname{vec}(E_a)^{T} \left( x_0 \otimes  C_{j}\right).
\end{equation}
Define
\begin{equation}
	Y_j\triangleq\left[Y_{1j}, Y_{2j}, \cdots, Y_{Lj}\right]^T,
\end{equation}
and
\begin{equation}
	B\triangleq\left[\operatorname{vec}({E}_1), \operatorname{vec}({E}_2), \cdots, \operatorname{vec}({E}_L)\right]^T.
\end{equation}
Thus we have
\begin{equation}\label{eq30}
	B\left( x_0\otimes  C_j\right)=Y_{j}.
\end{equation}
Hence, when the processes employed are all generalized-unital, the problem of identifying the quantum state and detector simultaneously can  be formulated as
\begin{problem}\label{problem1}
	Given the matrix $ B$ and experimental data $\hat Y_j$ for $1\leq j \leq M$, solve $\min_{x_0,\{C_j\} }\sum_{j=1}^{M}||\hat Y_j-B\big(x_0\otimes C_j\big)||^2$ where $x_0$ is the parameterization of $\rho_0$ and $C_j$ is the main part of the parameterization of $P_j$ for $1\leq j \leq M$, both under the basis $\{\Omega_j\}_{j=1}^{d^2-1}$.
\end{problem}

To obtain $\hat Y_{j}$ in Problem \ref{problem1}, we require the knowledge of $\{x_{a,0}\}_{a=1}^{L}$ and $\{C_{j,0}\}_{j=1}^{M}$, which cannot be directly estimated using Problem \ref{problem1}. If the $a$-th quantum process employed is TP, then $\operatorname{Tr}(\rho_a)=1$, and $x_{a,0}=\frac{1}{\sqrt{d}}$. Otherwise, if $\mathcal{E}_a$ is non-TP process, we can apply the measurement operator $P=I$, which is relatively straightforward to generate in an experimental setting, on the output state $\rho_a$ with $N_0$ copies, yielding an estimate $\hat x_{a,0}$. Similarly, to ascertain the trace of the detector, $\sqrt{d}C_{j,0}$, we employ a maximally mixed state $\rho=\frac{I}{d}$, which is also relatively straightforward. Subsequently, we apply the detector $\{P_j\}_{j=1}^{M}$ to measure this maximally mixed state with $N_0$ copies, resulting in the observation $\hat C_{j,0}$. Consequently, for non-TP processes, we have $\hat Y_{aj} =\hat{y}_{aj}- \hat x_{a,0} \hat C_{j,0}$, while for TP processes, $\hat Y_{aj} =\hat{y}_{aj}-  \hat C_{j,0}/\sqrt{d}$. In this way, we can obtain the value of $\hat Y_j=\left[\hat Y_{1j}, \hat Y_{2j}, \cdots, \hat Y_{Lj}\right]^T$.

\subsection{Second version of problem formulation} 
To extend the framework to arbitrary quantum processes, we propose the second version of problem formulation based on the natural basis $\{|l\rangle\langle k|\}_{1\leq l,k\leq d}$ \cite{8022944}.

Using \eqref{property2} and \eqref{eq8}, we have
\begin{equation}
	\begin{aligned}
		y_{aj}&=\operatorname{Tr}\left(P_j \rho_a\right)=\operatorname{vec}\left(P_{j}\right)^{\dagger} \operatorname{vec}\left(\rho_a\right)\\
		&=\operatorname{vec}\left(P_{j}\right)^{\dagger} \Big(\sum_{i=1}^{d^2}\left(A_i^a\right)^* \otimes A_i^a\Big) \operatorname{vec}\left(\rho_0\right)\\
		&=\left(\operatorname{vec}\left(\rho_0\right)^T \otimes \operatorname{vec}\left(P_{j}\right)^{\dagger}\right) \operatorname{vec}\Big(\sum_{i=1}^{d^2}\left(A_i^a\right)^* \otimes A_i^a\Big) \\
		& =\Big(\operatorname{vec}\Big(\sum_{i=1}^{d^2}\left(A_i^a\right)^* \otimes A_i^a\Big)\Big)^T\left(\operatorname{vec}\left(\rho_0\right) \otimes \operatorname{vec}\left(P_{j}^{T}\right)\right),
	\end{aligned}
\end{equation}
where the third line comes from \eqref{property2}.
Define 
\begin{equation}\label{ba}
	y_j\triangleq [y_{1j}, y_{2j}, \cdots, y_{Lj}],\;\; \mathcal{B}_a\triangleq\sum_{i=1}^{d^2}\left(A_i^a\right)^* \otimes A_i^a,
\end{equation}
and matrix $\mathcal{B}$ as
\begin{equation}
	\mathcal{B}\triangleq\left[\operatorname{vec}\left(\mathcal{B}_1\right), \cdots,\operatorname{vec}\left(\mathcal{B}_L\right)\right]^{T}.
\end{equation}
Thus, we have
\begin{equation}
	\mathcal{B}\left(\operatorname{vec}\left(\rho_0\right) \otimes \operatorname{vec}\left(P_{j}^T\right)\right)=y_{j}.
\end{equation}
The second version problem to identify the quantum state and detector simultaneously can thus be formulated as follows.
\begin{problem}\label{problem11}
	Given the matrix $ \mathcal{B}$ and experimental data $\hat y_{j}$ for  $1\leq j \leq M$, solve $$\min_{\rho_0,\{P_j\} }\sum_{j=1}^{M}||\hat y_{j}-\mathcal{B}\big(\operatorname{vec}(\rho_0) \otimes \operatorname{vec}(P_{j}^T)\big)||^2$$ where $\rho_0\in \mathcal{K}$ and $\{P_j\}_{j=1}^{M} \in \mathcal{R}$.
\end{problem}

 Problem \ref{problem11} can be reformulated as  Problem \ref{problem1}  if $\{x_{a,0}\}_{a=1}^{L}$ and  $\{C_{j,0}\}_{j=1}^{M}$ are known and the processes are generalized-unital.

\subsection{Comments on the two versions and definitions of informationally completeness/incompleteness }

Problem \ref{problem1} includes Hermitian constraints on the quantum state and detector, and the unit trace constraint on the state within the cost function, making it suitable for optimization in $\mathbb{R}$.
However, Problem \ref{problem1} is restricted to generalized-unital processes. On the other hand, Problem \ref{problem11} operates for arbitrary processes in $\mathbb{C}$ and does not need to consider $\{x_{a,0}\}_{a=1}^{L}$ and  $\{C_{j,0}\}_{j=1}^{M}$ separately, making it more appropriate for  pure input states as discussed in Section \ref{secexample}. The choice from these two
 formulations relies on factors such as the specific constraints and properties of the problem, as well as computational considerations.

Solving Problem \ref{problem1} or \ref{problem11} does not necessitate assuming a known state or detector. Hence, even in the presence of SPAM errors, as long as the known quantum processes $\{\mathcal{E}_a\}_{a=1}^{L}$ are accurate, one can reliably estimate the actual state and detector. The estimation results already incorporate the effect of SPAM errors, regardless of their strength. A limitation of our method is its dependence on the precise implementation of known quantum processes. Nevertheless, several high-accuracy preparation and identification algorithms for quantum processes have been discussed in the literature \cite{zhang1,zhang2,Gebhart2023}. The randomized benchmarking method \cite{PhysRevA.77.012307,PRXQuantum.3.020357} allows for the calibration of quantum processes or quantum gates independently of SPAM errors. This enables the accurate characterization of quantum gates, which can subsequently be utilized for tasks such as simultaneous identification of quantum states and detectors.

 Previous work in \cite{PhysRevA.98.042318} is restricted to unitary processes and requires the preparation of several probe states to apply QDT first, followed by QST. In qubit systems, Ref. \cite{PhysRevA.104.012416} does not require the preparation of probe states but is restricted to unitary processes and two-outcome POVMs. In contrast, our framework can be implemented for any dimensional quantum system and for generalized-unital or arbitrary processes, with the basis properly chosen.

We then provide the definition of an \emph{informationally complete} scenario as follows.
\begin{definition}\label{def1}
	A scenario is said to be \emph{informationally complete}  if
	$\operatorname{rank}(B)=(d^2-1)^2$ in Problem \ref{problem1} or $\operatorname{rank}(\mathcal{B})=d^4$ in Problem \ref{problem11}.
\end{definition}

Alternatively, $\operatorname{rank}(B)<(d^2-1)^2$  or $\operatorname{rank}(\mathcal{B})<d^4$ is referred to as the \emph{informationally incomplete} scenario.
The rank of $\mathcal{B}$ can be further characterized.
Assuming that there are $f$ groups of processes and in the $j$-th group, the Kraus operators of the $a$-th process are $\{A_i^{a,(j)}\}_{i=1}^{d^2}$. Define 
\begin{equation}
	\mathcal{A}_a^{(j)}\triangleq\sum_{i=1}^{d^2} ( A_{i}^{a,(j)})^{\dagger}  A_{i}^{a,(j)}.
\end{equation}
All the processes are grouped such that $ \mathcal{A}_1^{(j)}=\cdots=\mathcal{A}_{L_j}^{(j)}$ in the $j$-th group for all $1\leq j \leq f$.
 Thus, we totally have $L=\sum_{j=1}^{f}L_j$  processes. We propose the following theorem to characterize an upper bound on $\operatorname{rank}(\mathcal{B})$ and the proof is presented in Appendix \ref{appc}.
\begin{theorem}\label{th2}
We have	$$\operatorname{rank}(\mathcal{B})\leq \min \big(\sum_{j=1}^{f} \min \left(L_j, d^4-d^2+1\right), d^4\big).$$
\end{theorem}

Using Theorem \ref{th2}, we have presented the following corollary for TP processes where there is one group of the processes, i.e., $f=1$ and $\mathcal{A}_1^{(1)}=I$ (denoted as $\mathcal{A}_1$).
\begin{corollary}\label{th1}
	If $\mathcal{A}_1=\cdots=\mathcal{A}_L=I$,  we have	$\operatorname{rank}(\mathcal{B})\leq d^4-d^2+1$.
\end{corollary}

TP processes are quite common and widely implemented in quantum information processing \cite{qci}.
Using Corollary \ref{th1}, even if we prepare multiple TP quantum processes such that $L\geq d^4$, $\mathcal{B}$ is still rank-deficient. Therefore, based on Problem \ref{problem11}, TP processes are always informationally incomplete. However, $(d^2-1)^2< d^4-d^2+1$ means that proper TP processes can be informationally complete based on Problem \ref{problem1}. Hence, in the following, we firstly consider to solve Problem \ref{problem1}.

\section{Closed-form algorithm}\label{closed}
After obtaining all the measurement results, in this section, we design a closed-form algorithm for Problem \ref{problem1}, followed by another slightly modified algorithm for Problem \ref{problem11}. Then we
analyze the corresponding MSE scalings for both QST and QDT.

\subsection{Algorithm design}\label{sec31}
We start from investigating Problem \ref{problem1}.
To obtain a closed-form solution, we split Problem \ref{problem1}  into two sub-problems.

\addtocounter{problem}{-2}
\renewcommand{\theproblem}{\arabic{problem}{.1}}
\begin{problem}\label{subproblem1}
	Given the matrix $ B$ and experimental data $\hat Y_j$, solve $\min_{\{z_j\}}\sum_{j=1}^{M}||\hat Y_j-Bz_j||^2$ where $z_j \in \mathbb{R}^{(d^2-1)^2}$ for $1\leq j \leq M$.
\end{problem}

\renewcommand{\theproblem}{\arabic{problem}}
\addtocounter{problem}{-1}
\renewcommand{\theproblem}{\arabic{problem}{.2}}
\begin{problem}\label{subproblem2}
	Given $\hat z_j \in \mathbb{R}^{(d^2-1)^2} (1\leq j \leq M)$, solve $\min _{\tilde x_0, \{\tilde C_j \} } \sum_{j=1}^M \|\hat z_j-\tilde x_0 \otimes \tilde C_j\|^2$  where $\tilde x_0 \in \mathbb{R}^{d^2-1}$ is the parameterization of $ \rho_0$ and $\tilde C_j\in \mathbb{R}^{d^2-1}$ is the main part of the parameterization of $P_j$.
\end{problem}
\renewcommand{\theproblem}{\arabic{problem}}

For Problem \ref{subproblem1}, obviously we can minimize $||\hat Y_j-Bz_j||^2$ among $z_j$ for each $j$ independently. When $B$ is full-rank, using the least squares method, we can obtain a unique optimal estimate $\hat z_j$ as
\begin{equation}\label{s1}
\hat z_j=(B^TB)^{-1}B^{T}\hat Y_j
\end{equation}
for $1\leq j \leq M$.
When $B$ is rank-deficient, we can reconstruct $\hat z_j$ as 
\begin{equation}\label{ss1}
	\hat z_j=B^{+}\hat{Y}_j,
\end{equation}
where $B^{+}$ is the Moore–Penrose (MP) inverse of $B$.
We also consider  adding a regularization term as $||\hat Y_j-Bz_j||^2+z_j^{T} D z_j$
where $D \geq 0$ is the regularization matrix. Using this technique, we can also obtain a closed-form estimate
\begin{equation}\label{ss2}
\hat z_j=(B^TB +D)^{-1}B^{T}\hat Y_j.
\end{equation}
The topic of designing the regularization matrix $D$ and corresponding hyperparameters in $D$ has been discussed in kernel-based regularization of system identification \cite{6883125,Chen2018,10273596}, and in QST \cite{MU2020108837} and QDT \cite{wang2019twostage}.

For each $j \in \{1,2, \cdots, M\} $,
we can define a permuted version of $\hat{z}_{j}$ as $\mathcal{R}(\hat{z}_{j}) $  \cite{LOAN200085} where 
\begin{equation}
	\mathcal{R}(\hat{z}_{j})=\!\!\begin{bmatrix}\left(z_j\right)_{1:d^2-1}^T\\\left(z_j\right)_{d^2:2(d^2-1)}^T\\\vdots\\\left(z_j\right)_{(d^2-2)(d^2-1)+1:(d^2-1)^2}^T\end{bmatrix} \!\in \mathbb{R}^{(d^2-1)^2 \times (d^2-1)^2},
\end{equation}
and
\begin{equation}\label{sq2}
\|\hat z_j-\tilde x_0 \otimes \tilde C_j\|=	\|	\mathcal{R}(\hat{z}_{j})- \tilde x_0 \tilde C_j^{T}\|.
\end{equation}
Thus, if $M=1$,
Problem \ref{subproblem2} is   a  nearest Kronecker product problem \cite{LOAN200085} which can be solved efficiently by the Singular Value Decomposition (SVD) because it is the nearest rank-$1$ matrix problem  \cite{LOAN200085}.
 In fact, if $\tilde x_0$ and $\tilde C_j$ are the solutions, $q\tilde x_0$ and $\frac{1}{q}\tilde C_j$ for arbitrary $q$ is also a solution.
Therefore, to determine $q$, we need to determine one parameter in $x_0$ or $C_j$. In practice, any non-trivial observable $O \;(O\neq cI$ for certain $c\in \mathbb{R} )$ on $\rho_0$ can be employed to estimate $q$. Therefore, we may choose a highly accurate observable based on the experimental setting. For example, we can measure  $x_{0,1}$ using $N_0$ copies and obtain $\bar x_{0,1}$. Moreover, as highlighted in \cite{PhysRevA.104.012416}, 
$q$ can also be determined based on the experimental setup.
 Then the unique solution is 
\begin{equation}\label{cc1}
	\bar x_0=\frac{\bar x_{0,1}}{\tilde{x}_{0,1}}\tilde{x}_0, \;\bar C_j=\frac{\tilde{x}_{0,1}}{\bar x_{0,1}}\tilde{C}_j.
\end{equation}
When $M>1$, we plan to use the above SVD method to solve $\min _{\tilde x_0, \tilde C_j} \|\hat z_j-\tilde x_0 \otimes \tilde C_j\|$ for each $j\in\{1,2,\cdots,M\}$.
We need to solve Problem \ref{subproblem2} $M$ times to obtain the estimate of each POVM element.
In addition, for each $j $, we can obtain a temporary estimate of $\tilde x_0$, denoted as $\bar x_{0}^{(j)}$. After using this SVD method $M$ times and obtaining all the estimates $\bar x_0^{(j)}, 1\leq j \leq M$, we can choose one of them or take the average $\frac{1}{M} \sum_{j=1}^{M}\bar x_0^{(j)}$
 as the final estimate of $\tilde{x}_0$. This will not affect the error analysis in the next subsection.

 Using $\bar{x}_0$ and $\{\bar C_j\}_{j=1}^{M}$, we can reconstruct 
\begin{equation}
	\begin{aligned}
\bar	\rho_0 =\frac{1}{\sqrt{d}}\Omega_{0}+\sum_{k=1}^{d^2-1} \bar x_{0,k}\Omega_k, \;
\bar	P_j =\hat C_{j,0} \Omega_0+\sum_{k=1}^{d^2-1} \bar C_{j,k} \Omega_k.
	\end{aligned}
\end{equation} 
The estimate $	\bar\rho_0$ may not satisfy the positive semidefinite constraint. Thus we implement the fast correction algorithm \cite{effqst} on its eigenvalues, and obtain the final estimate $\hat\rho_0$. For QDT, the estimate $\{\bar P_j\}_{j=1}^{M}$ may not satisfy the completeness and positive semidefinite constraints.  Thus we implement the Stage-2 algorithm in \cite{wang2019twostage} to satisfy these constraints, and obtain the final estimate $\{\hat{P}_j\}_{j=1}^{M}$. In addition, these correction algorithms on the quantum state and detector are also analytical. The total number of state copies is $N=(2L+2)N_0$ for non-TP processes, which are used to obtain $\{\hat y_{aj}\}$, $\{\hat x_{a,0}\}$, $\{\hat C_{j,0}\}$, and $\bar{x}_{0,1}$, and $N=(L+2)N_0$ for TP processes, used to obtain $\{\hat y_{aj}\}$, $\{\hat C_{j,0}\}$, and $\bar{x}_{0,1}$.

Now we consider to solve Problem \ref{problem11} using a similar closed-form solution. Note that we do not need to prepare maximally mixed state $ \rho=\frac{I}{d}$ and $P=I$ to obtain $\{\hat C_{j,0}\}_{j=1}^{M}$ and $\{\hat x_{a,0}\}_{a=1}^{L}$ anymore.
We  also split Problem \ref{problem11} into two sub-problems.
\addtocounter{problem}{0}
\renewcommand{\theproblem}{\arabic{problem}{.1}}
\begin{problem}\label{subproblem21}
	Given the matrix $ \mathcal{B}$ and experimental data $\hat y_j$, solve $\min_{\{z_j\}}\sum_{j=1}^{M}||\hat y_j-\mathcal{B}z_j||^2$ where $z_j \in \mathbb{C}^{d^4}$ for $1\leq j \leq M$.
\end{problem}
\renewcommand{\theproblem}{\arabic{problem}}
\addtocounter{problem}{-1}
\renewcommand{\theproblem}{\arabic{problem}{.2}}
\begin{problem}\label{subproblem22}
	Given $\hat z_j \in \mathbb{C}^{d^4}$, $1\leq j \leq M$,  solve $\min _{\tilde \rho_0, \{\tilde P_j\}} \sum_{j=1}^{M} \|\hat z_j-\operatorname{vec}(\tilde \rho_0) \otimes \operatorname{vec}(\tilde P_j^{T})\|^2$.
\end{problem}
\renewcommand{\theproblem}{\arabic{problem}}

The solution to Problem \ref{subproblem21} is similar to solving Problem \ref{subproblem1}. When $\operatorname{rank}(\mathcal{B})=d^4$, the unique optimal solution is 
\begin{equation}
	\hat z_j=(\mathcal{B}^{\dagger}\mathcal{B})^{-1}\mathcal{B}^{\dagger}\hat y_j.
\end{equation}
When $\mathcal{B}$ is rank-deficient, we can also apply MP inverse or regularization.
For Problem \ref{subproblem22},   we can also utilize SVD to obtain  $\tilde \rho_0^{(j)}, \tilde P_j$ for each $j \in \{1,2, \cdots, M\} $. Since $\operatorname{Tr}(\rho_0)=1$, we do not need to measure the parameters in $x_0$ which is different from solving Problem \ref{subproblem2}.  After obtaining all the $\tilde \rho_0^{(j)}$, we take one $\tilde{\rho}_{0}^{(j)}$ or the average $\frac{1}{M}\sum_{j=1}^{M} \tilde{\rho}_0^{(j)}$ as $\tilde{\rho}_0$.
Then we need to correct it as a density operator. Therefore, we apply
\begin{eqnarray}\label{ad1}
	\tilde	\rho_0^{\prime}= \frac{\tilde	\rho_0+\tilde	\rho_0^{\dagger}}{2}, \;\bar	\rho_0=\frac{\tilde	\rho_0^{\prime}}{\operatorname{Tr}(\tilde	\rho_0^{\prime})}.
\end{eqnarray}
For each POVM element $\tilde{P}_j$, similarly we correct it as $\bar	P_j = (\tilde	P_j+\tilde	P_j^{\dagger})/2$.
Then we can apply the correction algorithms in \cite{effqst,wang2019twostage} to satisfy the positive semidefinite constraints, and obtain the final estimate $\hat{\rho}_0$, $\{\hat{P}_j\}_{j=1}^{M}$.

\begin{figure}
	\centering
	\includegraphics[width=3.6in]{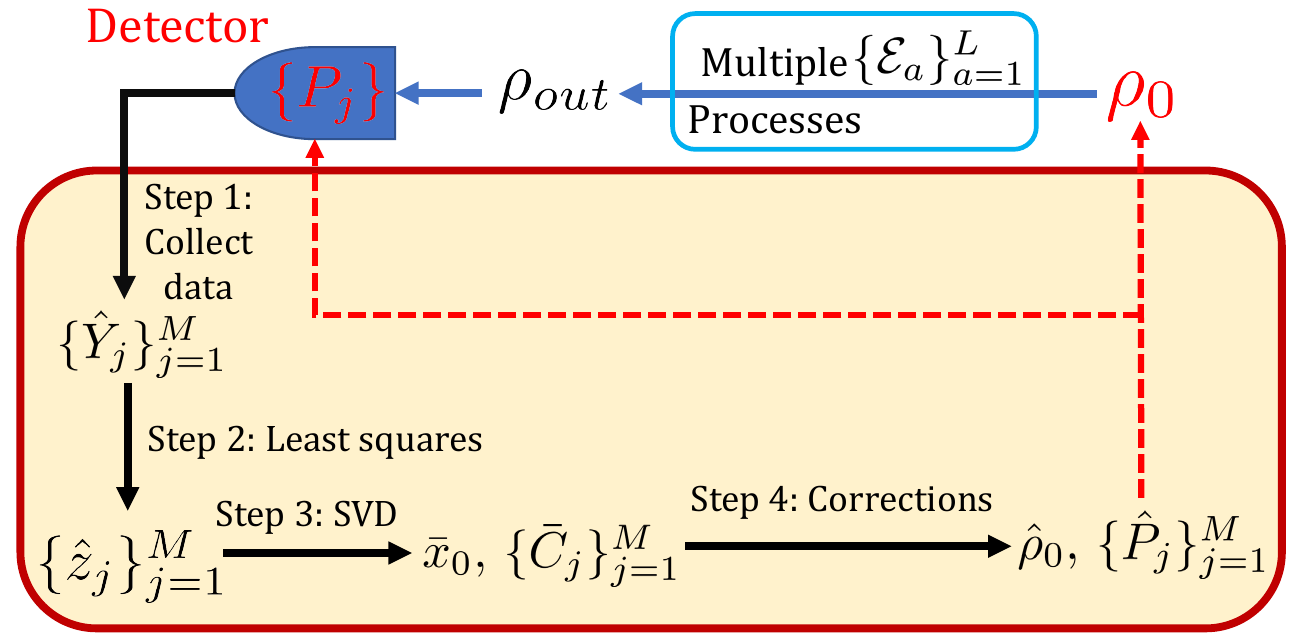}
	\centering{\caption{Procedures of our closed-form algorithm with four steps. Step 1 involves data collection. In Step 2, least squares is utilized to address Problem \ref{subproblem1} or Problem \ref{subproblem21}, and Step 3 employs SVD to tackle Problem \ref{subproblem2} or Problem \ref{subproblem22}. Finally, in Step 4,  the estimate is refined to ensure compliance with all the physical constraints.}\label{f2}}
\end{figure}

Overall,
the procedures of our closed-form algorithm have four steps as outlined in Fig. \ref{f2}.
Similar to \cite{wang2019twostage,8022944}, we can also derive the computational complexity for our closed-form solution. The total computational complexity is $O(MLd^4+Md^6)$, dominated by Steps 2 and 3.

\subsection{Error analysis}
Here we present the following theorem to analytically characterize the error scaling using our closed-form algorithm for Problem \ref{problem1}.
\begin{theorem}\label{t1}
In the informationally complete scenario,	 the MSE scalings in QST and QDT of our algorithm satisfy $\mathbb E||\hat \rho_0-\rho_0||^2=O\left({1}/{N}\right)$ and $\mathbb E\sum_{j=1}^{M}||\hat P_j-P_j||^2=O\left({1}/{N}\right)$ where $N$ is the number of state copies and $\mathbb E(\cdot)$ denotes the expectation with respect to all the possible measurement results.
\end{theorem}
\begin{IEEEproof}
	\subsubsection{Error in Step 1}
	Based on the analysis on the measurement results in Section \ref{sec2}, for TP processes, we have
	\begin{equation}\label{ny}
		\begin{aligned}
		\mathbb E|\hat y_{aj}-y_{aj}|^2=O\left(\frac{1}{N}\right), \; 	\mathbb E|\hat C_{j,0}-C_{j,0}|^2=O\left(\frac{1}{N}\right),
\end{aligned}
	\end{equation}
	and for non-TP processes, we also have
	\begin{equation}\label{ny2}
		\mathbb E|\hat x_{a,0}-x_{a,0}|^2=O\left(\frac{1}{N}\right).
	\end{equation}
Since
\begin{equation}
	\begin{aligned}
	&|\hat x_{a,0}\hat C_{j,0}-x_{a,0}C_{j,0}|\\
	=&	|\hat x_{a,0}\hat C_{j,0}- \hat x_{a,0} C_{j,0}+ \hat x_{a,0} C_{j,0}-x_{a,0}C_{j,0}|\\
	\leq& |\hat x_{a,0}||\hat C_{j,0}- C_{j,0}|+|C_{j,0}||\hat x_{a,0}-x_{a,0}|,
	\end{aligned}
\end{equation}	
also using \eqref{ny} and \eqref{ny2}, we have $\mathbb{E}|\hat x_{a,0}\hat C_{j,0}-x_{a,0}C_{j,0}|^2=O(1/N)$.
Since 
\begin{equation}
	|\hat Y_{aj}-Y_{aj}|=|(\hat y_{aj}-\hat x_{a,0}\hat C_{j,0})-(y_{aj}-x_{a,0}C_{j,0})|,
\end{equation}
we  also have
\begin{equation}\label{ey}
	\mathbb{E}|\hat Y_{aj}-Y_{aj}|^2=O\left(\frac{1}{N}\right), \; \mathbb{E}||\hat Y_{j}-Y_{j}||^2=O\left(\frac{1}{N}\right)
\end{equation}
for both non-TP and TP $(\hat x_{a,0}=x_{a,0}=1)$  processes.

	\subsubsection{Error in Step 2}
Using \eqref{ey}, we have 
	\begin{equation}
		\begin{aligned}
\mathbb E\left\|\hat{z}_j-z_j\right\|^2=&\mathbb E\left\|\hat{z}_j-{x}_0 \otimes {C}_{j}\right\|^2\\
=&	\frac{1}{N_0}\operatorname{Tr}\Big[\left(B^T B\right)^{-1}B^T R_{Y_j} B\big(B^T B\big)^{-1}\Big]\\
=&O\left(\frac{1}{N}\right),
\end{aligned}
		\end{equation}
where $R_{Y_j}$ is a constant matrix determined by the true measurement result $Y_j$ \cite{Qi2013}. 

	\subsubsection{Error in Step 3}
Since $\tilde x_0$ and $\tilde C_j$ minimize $\|\hat z_j-\tilde x_0 \otimes \tilde C_j\|$, we can obtain
	\begin{equation}
	\mathbb E	\left\|\hat{z}_j-\tilde{x}_0 \otimes \tilde{C}_j\right\|^2 \leq \mathbb E\left\|\hat{z}_j-x_0 \otimes C_{j}\right\|^2=O\left(\frac{1}{N}\right).
	\end{equation}
	Since 
	\begin{equation}
		\begin{aligned}
					&\left\|\tilde{x}_0 \otimes \tilde{C}_j-{x}_0 \otimes {C}_j\right\|\\
					\leq& \left\|\hat{z}_j-\tilde{x}_0 \otimes \tilde{C}_j\right\|+\left\|\hat{z}_j-x_0 \otimes C_{j}\right\|,
		\end{aligned}
	\end{equation}
	we have $\mathbb E \left\|\tilde{x}_0 \otimes \tilde{C}_j-{x}_0 \otimes {C}_j\right\|^2=O\left({1}/{N}\right)$.
	Therefore, using \eqref{cc1}, $\mathbb E \left\|\bar{x}_0 \otimes \bar{C}_j-{x}_0 \otimes {C}_j\right\|^2=O\left({1}/{N}\right)$
and	thus	the error in the first element also scales as
	\begin{equation}\label{first}
		\mathbb{E} \left(\bar{x}_{0,1}\bar{C}_{j,1}-x_{0,1}C_{j,1}\right)^2 =O\left(\frac{1}{N}\right).
	\end{equation}
	From the measurement process on the first element in  $x_0$ and obtain $\bar x_{0,1}$, we know
	\begin{equation}\label{first2}
		\mathbb E	\left\|\bar x_{0,1}-x_{0,1}\right\|^2 =O\left(\frac{1}{N}\right).
	\end{equation}
Since
\begin{equation}
\begin{aligned}
	&\left(x_{0,1}\bar{C}_{j,1}-x_{0,1}C_{j,1}\right)^2\\
	=&\left(x_{0,1}\bar{C}_{j,1}-\bar{x}_{0,1}\bar{C}_{j,1}+\bar{x}_{0,1}\bar{C}_{j,1}-x_{0,1}C_{j,1}\right)^2\\
\leq&2\bar{C}_{j,1}^2\left(x_{0,1}-\bar{x}_{0,1}\right)^2+2\left(\bar{x}_{0,1}\bar{C}_{j,1}-x_{0,1}C_{j,1}\right)^2,
\end{aligned}
\end{equation}
using \eqref{first} and \eqref{first2},
we have
\begin{equation}
	\mathbb E	\left\|\bar C_{j,1}-C_{j,1}\right\|^2 =O\left(\frac{1}{N}\right).
\end{equation}
Similarly, we can obtain that the error scalings of each element in $\bar{x}_0- x_0$ and $\bar{C}_j-{C}_j$ are all $O(1/N)$.  Therefore, whether we take the average of all the estimated states or choose one as the final estimation result, the following equations always hold as
	\begin{equation}
	\begin{aligned}
\mathbb E\left\|\bar \rho_0-\rho_0\right\|^2&=	\mathbb E\left\|\bar{x}_0-x_0\right\|^2=O\left(\frac{1}{N}\right),\\
\mathbb E\|\bar{P}_j-{P}_j\|^2&=\mathbb E\|\bar{C}_j-{C}_j\|^2+\mathbb E|\bar{C}_{j,0}-{C}_{j,0}|^2=O\left(\frac{1}{N}\right),
		\end{aligned}
	\end{equation}
for $1\leq j \leq M$.

	\subsubsection{Error in Step 4}
We implement the correction algorithms in \cite{effqst,wang2019twostage}, which have been proven to maintain the 
MSE scaling, i.e.,
\begin{equation}\label{r1}
	\begin{aligned}
\mathbb{E}\left\|\hat{\rho}_0-\bar{\rho}_0\right\|^2&=O(1/N),\\
\mathbb{E}\left\|\hat{P}_j- \bar{P}_j \right\|^2&=O(1/N).
	\end{aligned}
\end{equation}
Since
\begin{equation}\label{r2}
	\begin{aligned}
		||\hat \rho_0-\rho_0||\leq & ||\hat \rho_0-\bar\rho_0||+||\bar \rho_0-\rho_0||,\\
		||\hat P_j-P_j||\leq & ||\hat P_j-\bar P_j||+||\bar P_j-P_j||,
	\end{aligned}
	\end{equation}
 the final MSEs also scale as 
\begin{equation}
	\begin{aligned}
			\mathbb E||\hat \rho_0-\rho_0||^2=O\left(\frac{1}{N}\right),\; \mathbb E\sum_{j=1}^{M}||\hat P_j-P_j||^2=O\left(\frac{1}{N}\right).
	\end{aligned}
\end{equation}
\end{IEEEproof}

Moreover, if we implement the closed-form solution based on Problem \ref{problem11} in the informationally complete scenario, the final MSE scaling is still $O\left({1}/{N}\right)$.
The main difference is \eqref{ad1}. Since 
\begin{equation}
	\mathbb{E}\|\tilde{\rho}_{0} -\rho_{0}\|=O\left(\frac{1}{N}\right),
\end{equation}
we have 
\begin{equation}
	\mathbb{E}\|\tilde{\rho}_{0}^{\dagger} -\rho_{0}\|=O\left(\frac{1}{N}\right).
\end{equation}
Using \eqref{ad1}, we have 
\begin{equation}
	\|\tilde{\rho}_{0}^{\prime} -\rho_{0}\|\leq 	\frac{1}{2}\|\tilde{\rho}_{0} -\rho_{0}\|+	\frac{1}{2}\|\tilde{\rho}_{0}^{\dagger} -\rho_{0}\|,
\end{equation}
and thus $\mathbb{E}\|\tilde{\rho}_{0}^{\prime} -\rho_{0}\|^2=O\left({1}/{N}\right)$.
Using Lemma \ref{lemma1} in Appendix \ref{appendixa}, we have
\begin{equation}
	\mathbb{E}|\operatorname{Tr}(\tilde{\rho}_0^{\prime}) -1|^2 =O\left(\frac{1}{N}\right).
\end{equation}
Let $\delta\triangleq\operatorname{Tr}(\tilde{\rho}^{\prime}) -1 $ and since
\begin{equation}
\left\|\bar	\rho_0-\rho_0\right\|=\left\|\frac{\tilde{\rho}_0^{\prime}-\rho_{0}(1+\delta)}{\operatorname{Tr}(\tilde{\rho}_0^{\prime})}\right\|\leq \left\|\frac{\tilde{\rho}_0^{\prime}-\rho_{0}}{\operatorname{Tr}(\tilde{\rho}_0^{\prime})}\right\|+\left\|\frac{\rho_{0}\delta}{\operatorname{Tr}(\tilde{\rho}_0^{\prime})}\right\|,
\end{equation}
we have
\begin{equation}\label{ff1}
	\mathbb{E}\left\|\bar	\rho_0-\rho_0\right\|^2=O\left(\frac{1}{N}\right).
\end{equation}
Therefore the MSE scaling in \eqref{ad1} is  $O\left({1}/{N}\right)$. Using \eqref{r1} and \eqref{r2}, the final MSE scaling is also $\mathbb E||\hat \rho_0-\rho_0||^2= O\left({1}/{N}\right)$.
Similarly to QST, we can also prove that the final MSE scaling of QDT is  $	\mathbb E\sum_{j=1}^{M}||\hat P_j-P_j||^2=O\left({1}/{N}\right)$.

Overall, our closed-form algorithm has $O(1/N)$  MSE scalings for QST and QDT simultaneously in Theorem \ref{t1}, which achieves the same scalings as separate entities in QST \cite{Qi2013} and QDT \cite{wang2019twostage}.

\section{Sum of squares optimization}\label{sosop}
Testing whether a polynomial $g(x) $ is non-negative for all $x \in \mathbb{R}^n$ is NP-hard even when the degree of $g(x)$ is only $4$ \cite{8263706}. However, a more manageable sufficient condition for $g(x)$ to be nonnegative is for it to be a sum of squares (SOS) polynomial, which can be expressed as:
\begin{equation}
	g(x) = \sum_{i=1}^r f_i^2(x)
\end{equation}
where $\{f_i(x)\}_{i=1}^{r}$ are polynomials. Determining whether a polynomial is a sum of squares can be reformulated as solving semidefinite programming (SDP), a type of convex optimization problem for which efficient numerical solution methods exist~\cite{sostools}. Moreover, if the optimal value of the dual problem of the SDP equals the optimal value of the SDP, the strong duality holds \cite{Boyd2004Convex}, allowing us to determine the optimal value of $x$.

Given that the cost function $\min_{x_0,\{C_j\} }\sum_{j=1}^{M}||\hat Y_j-B(x_0\otimes C_j)||^2$ in Problem \ref{problem1} is a non-negative polynomial, we can employ SOS optimization techniques to address it. The task of deriving a lower bound for the global minimum of a polynomial function through SOS optimization was thoroughly explored in \cite{parrilo2001}.



In addition to the consideration of constraints on $\rho_0$ and $\{P_j\}_{j=1}^{M}$, it is essential to address the Hermitian constraint on $\rho_0$ and $\{P_j\}_{j=1}^{M}$, alongside the unit trace constraint on $\rho_0$. These constraints are effectively met by selecting the basis $\{\Omega_{j}\}_{j=0}^{d^{2}-1}$. Moreover, the completeness constraint on $\{P_j\}_{j=1}^{M}$ can be expressed as:
\begin{equation}
	\sum_{j=1}^{M} P_j=I \Leftrightarrow	\sum_{j=1}^M [C_{j,0}, C_{j}^{T}]=[\sqrt d,0,\cdots, 0].
\end{equation}
The positive semidefinite constraints of $\rho_0$ and $\{P_j\}_{j=1}^{M}$ are intricate and can be described by a semialgebraic set. Regarding quantum states, these constraints have been extensively addressed in the literature \cite{Kimura2003,1440563}. We can utilize the following lemma to delineate the physical set characterizing $x_0$.

\begin{lemma} (\cite{Kimura2003,1440563})
	Define $k_p(\rho)$, $p=2,\cdots,d$  recursively by
	\begin{equation}
		pk_p(\rho)=\sum_{f=1}^p(-1)^{f-1}\operatorname{Tr}(\rho^f)k_{p-f}(\rho)
	\end{equation}
	with $k_0=k_1=1$. Define the semialgebraic set 
	\begin{equation}
		\mathcal{K}\triangleq\{x_0\in\mathbb{R}^{d^2-1}: k_p(h(x_0))\geq0, p=2,\cdots, d\}.
	\end{equation}
	Then $h(\cdot)$ (defined after \eqref{x0}) is an isomorphism mapping between $\mathcal{K}$ and $\mathcal{P}$.
\end{lemma}

For each POVM element $P_j$, we can also normalize them to a density matrix and obtain a similar semialgebraic set. Hence, $P_j$ is positive semidefinite if and only if $$\frac{C_j}{\sqrt{d}C_{j,0}} \in \mathcal{K}.$$
Similar to the closed-form solution, the total number of copies is $N=(2L+2)N_0$ for non-TP processes to obtain $\{\hat y_{aj}\}$, $\{\hat x_{a,0}\}$, $\{\hat C_{j,0}\}$, and $\bar{x}_{0,1}$, and $N=(L+2)N_0$ for TP processes to obtain $\{\hat y_{aj}\}$, $\{\hat C_{j,0}\}$, and $\bar{x}_{0,1}$. Since $\sum_{j=1}^{M} \hat C_{j,0}=\sqrt{d}$,
 we propose to tackle Problem \ref{problem1} by solving the following optimization problem:
\begin{equation}\label{sos}
	\begin{aligned}
		\min&\;(-\gamma)\\
		\text{s.t. } &\sum_{j=1}^{M}||\hat Y_j-B\left(x_0\otimes C_j\right)||^2-\gamma \text{ is SOS,}\\
		&	\sum_{j=1}^M  C_{j}=[0,\cdots, 0],\\
		& x_0 \in \mathcal{K},   \frac{C_j}{\sqrt{d}\hat C_{j,0}} \in \mathcal{K},\; \forall  1\leq j \leq M,\\
		&x_{0,1}= \bar{x}_{0,1}.
	\end{aligned}
\end{equation}

The constrained polynomial optimization problem \eqref{sos} can be effectively tackled using the \texttt{findbound} function within SOSTOOLS \cite{sostools}. Note that the optimization problem yields the lower bound $\gamma$ of the cost function $\sum_{j=1}^{M}||\hat Y_j-B(x_0\otimes C_j)||^2$. Thus there may be instances where the \texttt{findbound} function fails to provide the values of the optimization variables $x_0$ and $\{C_j\}$ if the lower bound cannot be attained. Nevertheless, when the function does return values for the optimization variables, it signifies that these values achieve the lower bound. Consequently, this lower bound represents the minimum value of the cost function in Problem~\ref{problem1}. 

An intriguing open problem lies in determining the minimal number of distinct quantum processes necessary to obtain the values of $x_0$ and $\{C_j\}$ using SOS optimization. In our numerical example, we find that SOS optimization is capable of providing these values, even within the incomplete information scenario. Remarkably, even when completeness and positive semidefinite constraints remain inactive, the tensor structure inherent in the problem can effectively reduce the required number of processes, which will be present in Section \ref{sec61}. However,  closed-form solutions cannot fully exploit this advantageous structure, presenting a notable drawback.

Additionally, it is worth noting that the computational complexity associated with SOS optimization is notably high, often restricting its applicability to larger systems. While our numerical example showcases its efficacy in a one-qubit system, extending its applicability to a two-qubit system can entail a significant computational burden unless potential properties such as symmetry are explored to reduce the complexity.

\section{Illustrative examples}\label{secexample}
In this section, we delve into several illustrative examples to demonstrate the implementation of our framework. We begin by employing closed quantum systems and mixed-unitary processes, both of which are generalized-unital and can be addressed using Problem \ref{problem1}.
Alternatively, we assume prior information indicating that the input state is pure, a useful property in quantum technologies, simplifying the constraints in SOS optimization.  This approach is grounded in the formulation of Problem \ref{problem11}. Moreover, when both the input state and POVM elements are in low-rank, their tensor product, $\rho_0 \otimes P_j$, also retains a low-rank property because $\mathrm{rank}(\rho_0 \otimes P_j)=\mathrm{rank}{(\rho_0)} \times \mathrm{rank}{(P_j)}$. Compressed sensing methods as outlined in \cite{Flammia2012} can thus be leveraged in principle, which is presented in Appendix \ref{cs}.

\subsection{Closed quantum systems}
The closed quantum system model is a fundamental model
 in quantum physics whose dynamics are driven by the Hamiltonian.
Here we assume that there are $\mathcal{S}$ different known Hamiltonians $\{H_i\}_{i=1}^{\mathcal{S}}$ which are Hermitian and $ \operatorname{Tr}\left(H_i\right)=0 $  without loss of generality \cite{zhang1}. Using these Hamiltonians at different evolution times, we can generate multiple unitary quantum processes.

For each  $H_i$, the Liouville-von Neumann equation \cite{qci} is 
\begin{equation}\label{c1e4}
	\dot{\rho}^{i}(t)=-\text{i}[{H}_{i},\rho^{i}(t)], {\rho}^{i}(0)=\rho_0,
\end{equation}
which characterizes the dynamics of the closed quantum system.
Using \eqref{rho}-\eqref{x0}, \eqref{c1e4} and the methods outlined in \cite{zhang1, alicki}, the linear dynamical equation for closed quantum systems driven by the Hamiltonian $H_i$ is
\begin{equation}\label{c1}
	\dot{{x}}^{i}(t)={R}_{i} {x}^{i}(t),
\end{equation}
where $ R_{i}$ can be calculated from
 $H_i$ and $\{\Omega_j\}_{j=0}^{d^2-1}$ (the details are omitted here), and $ R_{i}=-R_{i}^{T} $ due to the antisymmetry of the structure constants of $\mathfrak{su}(d)$ \cite{zhang1}.
Using \eqref{cj} and \eqref{c1}, the dynamical equation of the closed quantum system driven by the Hamiltonian $ H_i $ is
\begin{equation}\label{cm}
	\left\{\begin{aligned}
		\dot{x}^{i}(t)&=R_i x^{i}(t),\; x^{i}(0)=x_0, \\
		y_j^{i}(t)&=C_j^{T} x^{i}(t),
	\end{aligned}
	\right.
\end{equation}
where $y_j^{i}(t)\triangleq \operatorname{Tr}\left(P_j\rho^{i}(t)\right)-\frac{C_{j,0}}{\sqrt{d}}$.
Therefore, at time $ t $, the measurement result $ y_i^{j}(t) $ is 
\begin{equation}\label{h1}
	y_j^{i}(t)=C_{j}^{T} \exp{(R_i t)} x_0.
\end{equation}

In the experiment, measurements often yield discrete outcomes, and it is natural to adapt the following framework as in \cite{Merkel2010,xiaoqst}. Assuming a sampling interval of $\Delta t$ and a total of $n$ different temporal sampling points, we utilize $N_0$ copies at each point $k\Delta t$. These copies undergo identical Hamiltonian evolution under $H_i$ over the duration $k\Delta t$. Subsequently, we apply the detector $\{P_j\}_{j=1}^{M}$ to measure the output state at time $k\Delta t$. Averaging the results from the $N_0$ copies provides $\hat y_j^{i}(k\Delta t)$, an estimate of the ideal value $y_j^{i}(k\Delta t)$. For simplicity, we denote $x^{i}(k \Delta t)$ and $y_j^{i}(k\Delta t)$ as $x^{i}(k)$ and $y_j^{i}(k)$, respectively. This entire process is repeated for each $k=1,2,\cdots,n$. Encompassing all $n$ sampling points, we define the \emph{time traces} as $\hat {\mathcal{Y}}^{i}_{j} \triangleq [\hat y_j^{i}(1), \cdots, \hat y_j^{i}(n)]$. The detailed measurement process can be found, for example, in \cite{Cole2005}. Transforming  the dynamic system equation \eqref{cm} into a discrete form, we have:
\begin{equation}\label{close2}
	\left\{
	\begin{aligned}
		x^{i}(k+1)&=Q_i x^{i}(k),\\
		y_j^{i}(k)&=C_j^{T} x^{i}(k),
	\end{aligned}
	\right.
\end{equation}
where $ Q_i=\exp \left(R_i \Delta t\right)$ and thus the measurement result is
\begin{equation}\label{dis1}
	y_j^{i}(k)=C_j^{T}(Q_i)^{k}x_0.
\end{equation}
Define the matrix $\mathcal{Q}_i$ as
\begin{equation}\label{q1}
	\mathcal{Q}_i\triangleq\left[\operatorname{vec}(\left(Q_i\right)^1), \operatorname{vec}(\left(Q_i\right)^2), \cdots,\operatorname{vec}(\left(Q_i\right)^n)\right],
\end{equation}
and ${Y}_j^{H}$ as
\begin{equation}
	{Y}_j^{H}\triangleq\left[\mathcal{Y}^{1}_{j},\mathcal{Y}^{2}_{j},\cdots,\mathcal{Y}^{L}_{j}\right]^T.
\end{equation}
Let $B^{H}$ be
\begin{equation}
	B^{H}\triangleq\left[\mathcal{Q}_1, \mathcal{Q}_2, \cdots, \mathcal{Q}_{\mathcal{S}}\right]^T,
\end{equation}
which is an $n\mathcal{S}\times (d^2-1)^2$ real matrix.
We then have
\begin{equation}\label{lq}
	B^{H}\left(x_0\otimes C_j\right)=Y_j^{H},
\end{equation}
for $1\leq j \leq M$, which has the same structure as \eqref{eq30}.
Therefore, we can also formulate it as an optimization problem as Problem \ref{problem1} and utilize the closed-form algorithm or SOS optimization to solve it.

To ensure the measurement data are informationally complete,
we propose the following proposition to characterize the minimum value of the type number of the Hamiltonians.
The proof is presented in Appendix \ref{ad}.
\begin{proposition}\label{p1}
	To ensure $B^{H}$ is full-rank, 
	at least $\frac{(d^2-1)^2}{d^2-d+1}$ different Hamiltonians are needed, i.e., $\mathcal{S}\geq \Big\lceil \frac{(d^2-1)^2}{d^2-d+1}\Big\rceil$.
\end{proposition}

In addition, to ensure $\operatorname{rank}(\mathcal{Q}_i)= d^2-d+1 $, the number of sampling points $n$ should be equal to or greater than $d^2-d+1$, i.e., $n\geq d^2-d+1$.

\vspace*{-8pt}

\subsection{Mixed-unitary quantum processes}
Here we consider a special quantum process: mixed-unitary quantum process as in \cite{Girard2022}
\begin{equation}
	\rho_{a}(t)=\sum_{i=1}^m \sigma_i^{a} U_i^{a}(t) \rho_0 (U_i^{a})^{\dagger}(t).
\end{equation}
where $\{U_i^{a}(t)\}_{i=1}^{m}$ are unitary operators.
If $ \sum_{i=1}^{m} \sigma_i^{a}=1, \sigma_i^{a}>0, \forall i$, the mixed-unitary  processes is TP and unital \cite{Girard2022}.
If $ \sum_{i=1}^{m} \sigma_i^{a}<1, \sigma_i^{a}>0, \forall i$, the mixed-unitary  processes is non-TP and generalized-unital.
For each unitary matrix  $U_i^{a}(t)$, let $U_i^{a}(t)=\exp{\left(-\mathrm{i}H_i^{a} t\right)}$ where $H_i^{a}$ is the Hamiltonian and can be calculated through Schur decomposition as presented in \cite{8022944}.
We can also construct an  antisymmetric matrix $R_i^{a}$ like \eqref{c1}. Therefore, the dynamics of the $a$-th mixed-unitary quantum process is
\begin{equation}\label{mixed}
	\left\{\begin{aligned}
		x^{a}(t)&=\sum_{i=1}^m \sigma_i^{a} \exp \left(R_i^{a} t\right) x_0,\\
		y_j^{a}(t)&=C_j^{T}x_{a}(t).
	\end{aligned}
	\right.
\end{equation}
Similar to closed quantum systems,
let the sampling time be $ \Delta t $ and we obtain $ n $ sampling points for each process. Define $ Q_i^{a}\triangleq\exp \left(R_i^{a} \Delta t\right)$ and we have
\begin{equation}\label{dis2}
	y_j^{a}(k)=C_j^{T}\sum_{i=1}^m \sigma_i^{a} (Q_i^{a})^{k} x_0,
\end{equation}
which has the same structure as \eqref{dis1} except changing $(Q_i)^{k}$ to a weighted summation. In the end, it is not difficult to arrive at an equation similar to \eqref{lq}.
Therefore, we can also formulate it into an optimization problem as in Problem \ref{problem1} and utilize our closed-form algorithm or SOS optimization to solve it.

\vspace*{-10pt}
\subsection{Pure input states}
Pure states serve as crucial quantum resources and find extensive application in experiments. Here, we assume that $\rho_0=|\psi\rangle\langle\psi|$ is a pure state.  Using \eqref{property2}, we have
\begin{equation}
	\operatorname{vec}(|\psi\rangle\langle\psi|)=|\psi\rangle^{*}\otimes |\psi\rangle.
\end{equation}
Hence, Problem \ref{problem11} can be converted into Problem \ref{problem31}.
\addtocounter{problem}{+1}
\begin{problem}\label{problem31}
	Given the matrix $ \mathcal{B}$ and experimental data $\hat y_{j}$ for  $1\leq j \leq M$, solve $\min_{|\psi\rangle,\{P_j\} }\sum_{j=1}^{M}||\hat y_{j}-\mathcal{B}\big(|\psi\rangle^{*}\otimes |\psi\rangle\otimes \operatorname{vec}(P_{j}^T)\big)||^2$ where $|\psi\rangle \in \mathbb{C}^{d}$ and $\{P_j\}_{j=1}^{M} \in \mathcal{R}$.
\end{problem}

To solve this problem with a closed-form solution, we can first implement the solution to Problem \ref{subproblem21} and Problem \ref{subproblem22} in Section \ref{sec31} and obtain $\bar \rho_0$ and $\{\bar P_j\}$. Assuming the spectral decomposition of $\bar \rho_0=\bar V \operatorname{diag}(\bar\lambda_1, \cdots, \bar\lambda_d) \bar V^{\dagger} $ where $\bar\lambda_1\geq \cdots \geq \bar\lambda_d$, the final estimate of the  pure input state is 
\begin{equation}\label{pure}
	\hat \rho_0=V \operatorname{diag}(1, 0, \cdots, 0) V^{\dagger}.
\end{equation}

For the error analysis, we already have
\begin{equation}
	\mathbb{E}	\|\bar \rho_0- |\psi\rangle\langle\psi| \|^2= O\left(\frac{1}{N}\right)
\end{equation}
from \eqref{ff1}.
Thus, using Lemma \ref{lemma1} in Appendix \ref{appendixa}, we have
\begin{equation}
	\mathbb{E} |\bar \lambda_1 -1 |^2=O\left(\frac{1}{N}\right), 	\mathbb{E} |\bar \lambda_i |^2=O\left(\frac{1}{N}\right), \forall\; 2\leq i \leq d.
\end{equation}
Hence,
\begin{equation}
		\mathbb{E}	\|\hat \rho_0- \bar \rho_0 \|^2= \mathbb{E} |\bar \lambda_1 -1 |^2+ \sum_{i=2}^{d}  \mathbb{E} |\bar \lambda_i |^2= O\left(\frac{1}{N}\right).
\end{equation}
Using \eqref{r2} again, the final MSE of the state tomography still scales as $O(1/N)$. Note that Ref. \cite{xiaorank} has proved that the infidelity $1-F(\hat{\rho}_0,\rho_0)$ also has the optimal scaling  $O(1/N)$ in this scenario.

Despite maintaining a closed-form algorithm that can be proven to have an $O(1/N)$ MSE scaling, it does not exploit the prior knowledge of pure input states, and only in its final part addresses the constraint of pure state through corrections. Alternatively, we can formulate the problem as an SOS optimization problem as follows:
\begin{equation}\label{sospure}
	\begin{aligned}
		\min&\;(-\gamma)\\
		\text{s.t. } &||\hat y_{j}-\mathcal{B}\big(|\psi\rangle^{*}\otimes |\psi\rangle\otimes \operatorname{vec}(P_{j}^T)\big)||^2-\gamma \text{ is SOS,}\\
		&\sum_{j=1}^M P_j=I, P_j \geq 0, \forall\; 1\leq j \leq M,\\
		& |\psi\rangle \in \mathbb{C}^{d}, \||\psi\rangle\|=1.\\
	\end{aligned}
\end{equation}
By using standard SOS solution tools in
this formulation, the pure state prior information is utilized throughout the whole solution process, obviating the need to consider the positive semidefinite constraint on quantum states. In particular, this problem is defined within the complex domain, necessitating a transformation to the real domain before employing SOSTOOLS for resolution.

\begin{remark}
	If we additionally possess prior knowledge that the measurement operator is projective, i.e., $P_j=|\phi_j\rangle\langle\phi_j|$, and utilize the closed-form solution, we can also refine the eigenvalues of $\{\bar P_j\}$ as described in \eqref{pure} and the MSE still scales as $O(1/N)$. Furthermore, we can frame this within the framework of an SOS optimization problem, where the constraints become:
	\begin{equation}
		\begin{aligned}
		&\sum_{j=1}^{M} ||\hat y_{j}-\mathcal{B}\big(|\psi\rangle^{*}\otimes |\psi\rangle\otimes |\phi_j\rangle^{*}\otimes |\phi_j\rangle\big)||^2-\gamma \text{ is SOS,}\\
		&|\psi\rangle, |\phi_j\rangle \in \mathbb{C}^{d}, \||\psi\rangle\|=1, \||\phi_j\rangle\|=1, \forall\; 1\leq j \leq M.\\
				\end{aligned}
	\end{equation}
As a result, there is no longer a need to consider the positive semidefinite constraint on the POVM elements. 
\end{remark}

\section{Numerical examples}\label{sec6}
In this section, we consider three numerical examples: one-qubit closed quantum systems, random one-qubit quantum processes with   pure input states and two-qubit mixed-unitary quantum processes.
The simulation is run on a laptop with i9-13980HX and  64G DDR5 memory size. For each data point in the figures of this section, we repeat our algorithm $50$ times and calculate the mean to obtain the MSE and error bar.
\subsection{One-qubit closed quantum systems}\label{sec61}
Here we consider a one-qubit example and prepare five Hamiltonians ($\mathcal{S}=5$)  as
\begin{equation}\label{hami}
	\frac{\sigma_x \pm \sigma_{\mathrm{y}}}{2}, \frac{\sigma_y \pm \sigma_z}{2}, \frac{\sigma_z + \sigma_x}{2}.
\end{equation}
 The unit of the Hamiltonians is MHz and the sampling time is $\Delta t=1\mu s$. The total number of sampling points for each Hamiltonian is $n=2^2-2+1=3$. 
Let the unknown initial quantum state be
\begin{equation}
	\rho_0=V\operatorname{diag}(0.1,0.9)V^\dagger,
\end{equation}
and the three-valued detector $(M=3)$ be
\begin{equation}\label{onede}
	\begin{aligned}
		P_{1}&=U_{1}\operatorname{diag}\big(0.4,0.1\big)U_{1}^{\dagger}, \\
		P_{2}&=U_{2}\operatorname{diag}\big(0.5,0.1\big)U_{2}^{\dagger}, \\
		P_{3}&=I-P_{1}-P_{2}\geq 0,
\end{aligned}\end{equation}
where $V$, $ U_1 $ and $ U_2 $ are random unitary matrices generated by the algorithms in \cite{MISZCZAK2012118,qetlab}.
We measure $\sigma_{x}$ on $\rho_0$ to determine $\bar x_{0,1}$.

For this one-qubit example of SOS optimization, $x_0 \in \mathcal{K}$ is equivalent to $x_{0,1}^2+x_{0,2}^2+x_{0,3}^2 \leq \frac{1}{2}$, coinciding with the Bloch sphere representation.
Therefore,
\eqref{sos}  can be expressed as 
\begin{equation}\label{sosone}
	\begin{aligned}
		\min  &\;-\gamma\\
		\text{s.t. } & \sum_{j=1}^{3}||\hat Y_j-B\left(x_0\otimes C_j\right)||^2-\gamma \text{ is SOS,}\\
		&\sum_{j=1}^3 C_j=[0, 0, 0],\\
		& \bar x_{0,1}^2+x_{0,2}^2+x_{0,3}^2 \leq \frac{1}{2},\\
		& C_{j,1}^2+C_{j,2}^2+C_{j,3}^2\leq\frac{d}{2}\hat{C}_{j,0}^2, \forall\; 1\leq j \leq 3,
	\end{aligned}
\end{equation}
which is solved by \texttt{findbound} function within SOSTOOLS. We find that in this case, {\ttfamily{findbound}}  can always output values of optimization variables and thus the lower bound $\gamma$ is the minimum value of the cost function. In addition, we check that the positive semidefinite constraints of state and POVM elements are all inactive in the optimal value, satisfying the analysis in Section \ref{sosop}.

\begin{figure}[htbp]
	\begin{minipage}[t]{0.5\linewidth}
		\centering
		\includegraphics[width=1\textwidth]{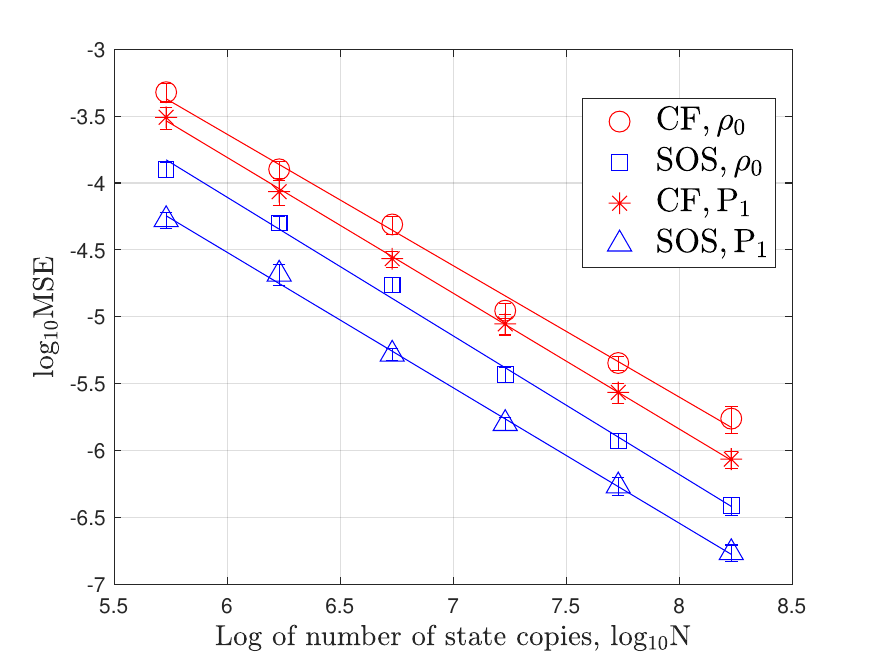}
	\end{minipage}\\
	\begin{minipage}[t]{0.5\linewidth}
		\centering
		\includegraphics[width=1\textwidth]{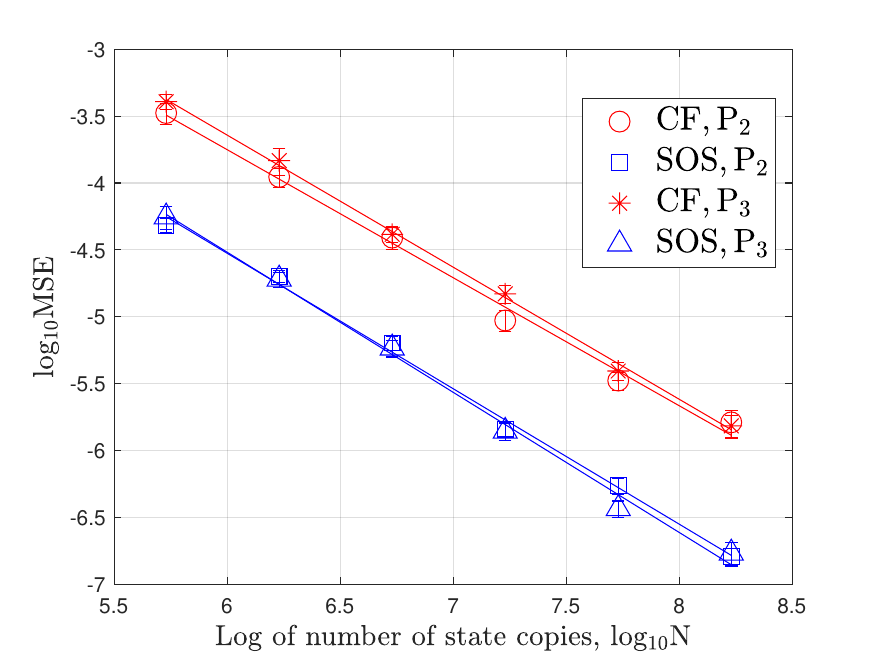}
		\end{minipage}
		\caption{Log-log plot of MSE versus  the total number of state copies $N$  for one-qubit closed quantum systems using the closed-form (CF) solution  and SOS optimization when the number of different Hamiltonians is $\mathcal{S}=5$.} \label{ff2}
\end{figure}

We compare the results of the closed-form (CF) solution in Section \ref{closed} and SOS optimization in Section \ref{sosop}.
The simulated estimation results are presented in Fig. \ref{ff2} where the MSE scalings of the quantum state and detector are both $O(1/N)$ using these two algorithms. In addition, the MSEs of SOS optimization are smaller than those of the closed-form solution.

%
%
%

\begin{figure}
	\centering
	\subfigure{
		\begin{minipage}[b]{0.5\textwidth}
			\centering	\includegraphics[width=1\textwidth]{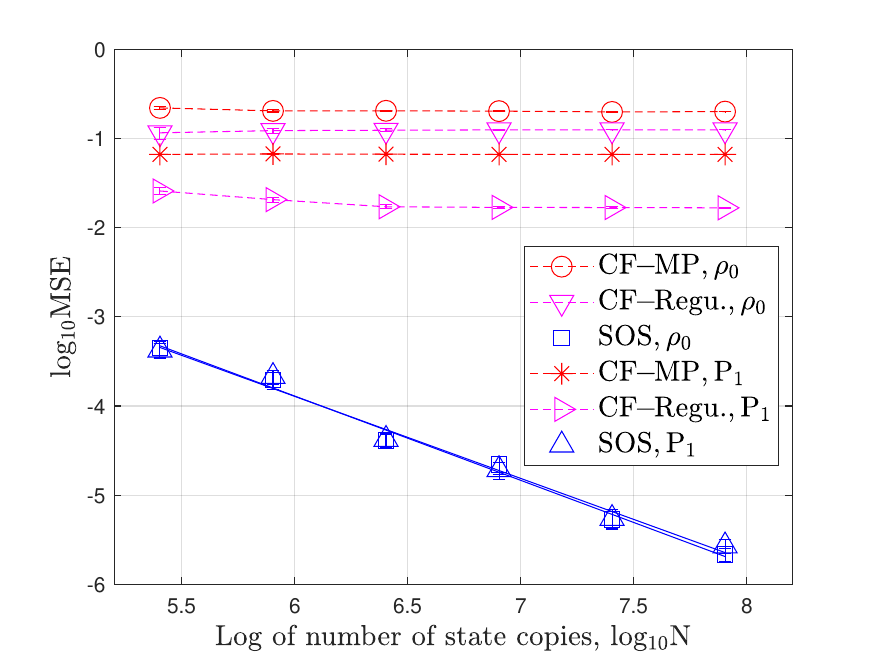}
		\end{minipage}
	}
	\subfigure{
		\begin{minipage}[b]{0.5\textwidth}
			\centering	\includegraphics[width=1\textwidth]{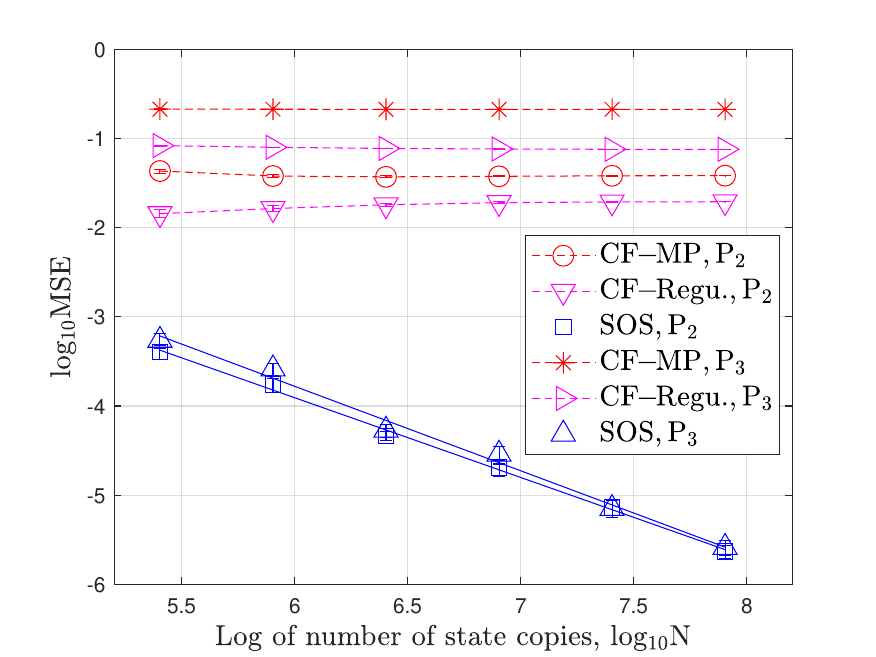}
		\end{minipage}
	}
	\caption{Log-log plot of MSE versus  the total number of state copies $N$ for one-qubit closed quantum systems using the closed-form (CF) solution,  Moore–Penrose (MP) inverse  and  regularization (Regu.), and SOS optimization when the number of different Hamiltonians is $\mathcal{S}=3$.} \label{f3}
\end{figure}

We then only utilize the first three Hamiltonians in \eqref{hami}   and the number of sample points is two ($n=2$), which is an informationally incomplete scenario. For the closed-form solution, we utilize \eqref{ss1} to obtain a unique 
MP inverse solution and \eqref{ss2} to obtain a unique 
regularization solution where we choose $D=\frac{100}{N}$.
   Other steps are the same as Section \ref{closed}. For SOS optimization, we also utilize SOSTOOLS to solve \eqref{sosone}.  In this case, {\ttfamily{findbound}}  can also always output  values of optimization variables and the minimum value of the cost function. In addition, the positive semidefinite constraints of state and POVM elements are all inactive in the optimal value. 
The results are presented in Fig. \ref{f3} where even as $B$ is rank-deficient, the MSE scalings of  the quantum state and detector are both $O(1/N)$ using SOS optimization. While using the closed-form solution with MP inverse and regularization, the MSEs of  the quantum state and detector are basically unchanged, because this specific $B$ is already informationally incomplete for Problem \ref{subproblem1} and thus the estimates $\hat z_j$ using \eqref{ss1} are farther away from the true value
compared with the previous simulation of the informationally complete scenario.

Table \ref{table1} presents the time consumption results for both the closed-form solution and SOS optimization. The closed-form solution shows significantly lower time consumption compared with SOS optimization. Although the closed-form solution is fast, it often compromises in accuracy. Conversely, SOS optimization demonstrates superior accuracy at the cost of considerably longer computational time. Hence, there is a trade-off between accuracy and time consumption between these two algorithms.
Additionally, besides ensuring the informational completeness of the data, the closed-form solution requires numerous distinct Hamiltonians, as indicated in Proposition \ref{p1}. However, SOS optimization can achieve a high accuracy with a reduced number of Hamiltonians.

\begin{table}
	\caption{Time consumption  of the closed-form solution in Section \ref{closed} and SOS optimization in  \eqref{sosone}.}
	\renewcommand{\arraystretch}{1.4}
	\label{table1} 
	
	\centering
	
	\begin{tabular}{|p{2.9cm}<{\centering}|p{4.3cm}<{\centering}|p{4cm}<{\centering}|}
		\toprule
		Setting&Closed-form solution &SOS\\
		\hline
		$L=5,n=3$ & $0.672$ sec &$ 1256.574 $ sec \\
		\hline
		$L=3,n=2$& $0.381$ sec  & $1214.843   $ sec\\
		\bottomrule
	\end{tabular}
	
\end{table}

\subsection{Random one-qubit quantum processes with a   pure input state}
\begin{figure}
	\centering
	\subfigure{
		\begin{minipage}[b]{0.5\textwidth}
			\centering\includegraphics[width=1\textwidth]{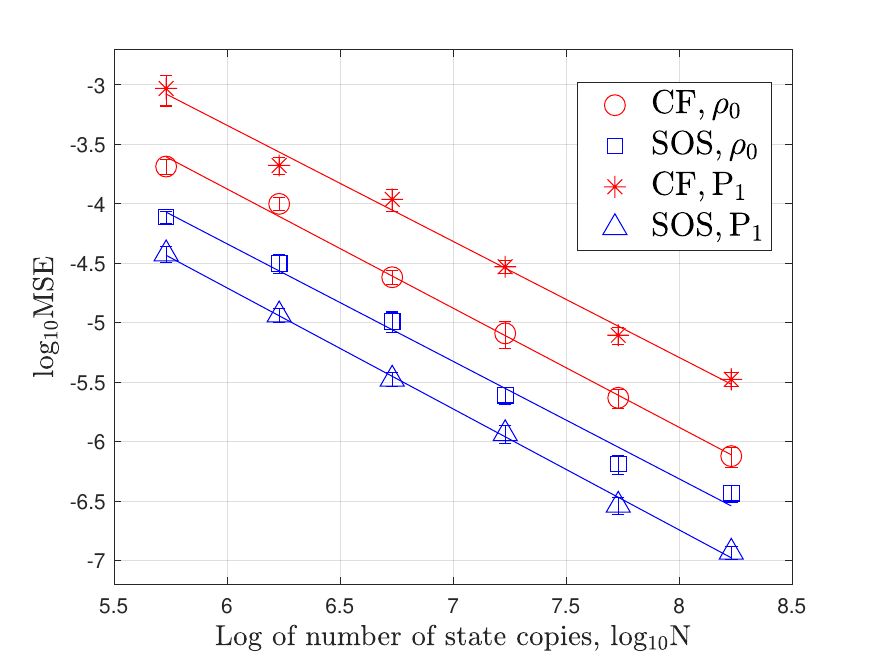}
		\end{minipage}
	}
	\subfigure{
		\begin{minipage}[b]{0.5\textwidth}
			\centering\includegraphics[width=1\textwidth]{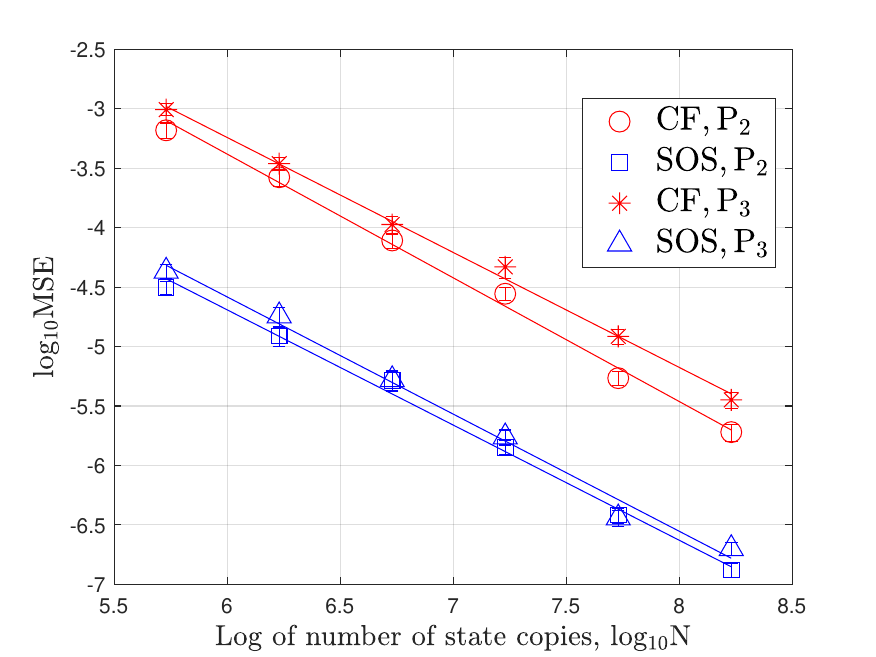}
		\end{minipage}
	}
	\caption{Log-log plot of MSE versus the total number of copies $N$  using the closed-form (CF) solution and SOS optimization for random one-qubit quantum processes with the pure input state.} \label{rand}
\end{figure}

Let the unknown input state be
\begin{equation}
	\rho_0=V\operatorname{diag}(1,0)V^\dagger,
\end{equation}
and the detector remains consistent with \eqref{onede} using random unitaries $V, U_1, U_2$. Initially, we generate $17$ non-TP quantum processes using the algorithm in \cite{qetlab}, which is informationally complete for Problem \ref{problem11}. Subsequently, we address the problem using both the closed-form solution and SOS optimization as outlined in \eqref{sospure}. The results are depicted in Fig. \ref{rand}, demonstrating that all the MSEs decrease at a rate of $O(1/N)$, with the MSE obtained through SOS optimization smaller than that achieved via the closed-form solution.

\begin{remark}
Bit flip and phase flip channels represent fundamental quantum operations on single qubits \cite{qci}. A bit flip channel can be characterized by two Kraus operators:
\begin{equation}
	A_1^{a}=\sqrt{p}\left[\begin{array}{cc}{1}&{0}\\{0}&{1}\end{array}\right], \; A_2^{a}=\sqrt{1-p}\left[\begin{array}{cc}0&1\\1&0\end{array}\right],
\end{equation}
where $p$ denotes the probability of the qubit flipping from $|1\rangle$ to $|0\rangle$. Generating multiple bit-flip channels involves varying $p$. However, both $\mathcal{B}$ in Problem \ref{problem11} and $B$ in Problem \ref{problem1} consistently exhibit rank deficiency ($\operatorname{rank}(\mathcal{B})=\operatorname{rank}(B)=2$), rendering them informationally incomplete.

Similarly, phase flip channels are characterized by Kraus operators:
\begin{equation}
	A_1^{a}=\sqrt{p}\left[\begin{array}{cc}{1}&{0}\\{0}&{1}\end{array}\right], \; A_2^{a}=\sqrt{1-p}\left[\begin{array}{cc}1&0\\0&-1\end{array}\right].
\end{equation}
It can be verified that both $\mathcal{B}$ and $B$ maintain rank deficiency ($\operatorname{rank}(\mathcal{B})=\operatorname{rank}(B)=2$).
\end{remark}

\subsection{Two-qubit mixed-unitary quantum processes}
Let the unknown initial quantum state be
\begin{equation}
	\rho_0=V\operatorname{diag}(0.1,0.2,0.3,0.4)V^\dagger,
\end{equation}
and the three-valued detector $(M=3)$ be
\begin{equation}\label{twodetector}
	\begin{aligned}
		P_{1}&=U_{1}\operatorname{diag}\big(0.1,0.1,0.1,0.3\big)U_{1}^{\dagger}, \\
		P_{2}&=U_{2}\operatorname{diag}\big(0.1,0.1,0.1,0.5\big)U_{2}^{\dagger}, \\
		P_{3}&=I-P_{1}-P_{2}\geq 0,
\end{aligned}\end{equation}
where $V$, $ U_1 $ and $ U_2 $ are random unitary matrices generated by the algorithms in \cite{MISZCZAK2012118,qetlab}.
We measure $\sigma_{x}$ on $\rho_0$ to determine $\bar x_{0,1}$.

For the mixed-unitary quantum process described in \eqref{mixed}, we set $ m=2 $ and generate two random Hamiltonians, $H_1$ and $H_2$, using algorithms from \cite{MISZCZAK2012118,qetlab}, with $\sigma_1^{a}=0.3$ and $\sigma_2^{a}=0.7$, $\forall a=1,2$. Consequently, the dynamics of each system can be expressed as
\begin{equation}
	\left\{\begin{aligned}
		x^{a}(t)&=0.3\exp \left(A_1^{a} t\right) x_0+0.7 \exp \left(A_2^{a} t\right) x_0, \\
		y^{a}_{j}(t)&=C_{j}^{T}x^{a}(t).
	\end{aligned}
	\right.
\end{equation}
where $A_1^{a}$ and $A_2^{a}$ are defined using \eqref{mixed}. We randomly generate a total of $30$ different pairs of Hamiltonians, $H_1$ and $H_2$, for the mixed-unitary quantum processes and the number of sampling points is $n=30$.

\begin{figure}
	\centering
	\subfigure{
		\begin{minipage}[b]{0.5\textwidth}
			\centering\includegraphics[width=1\textwidth]{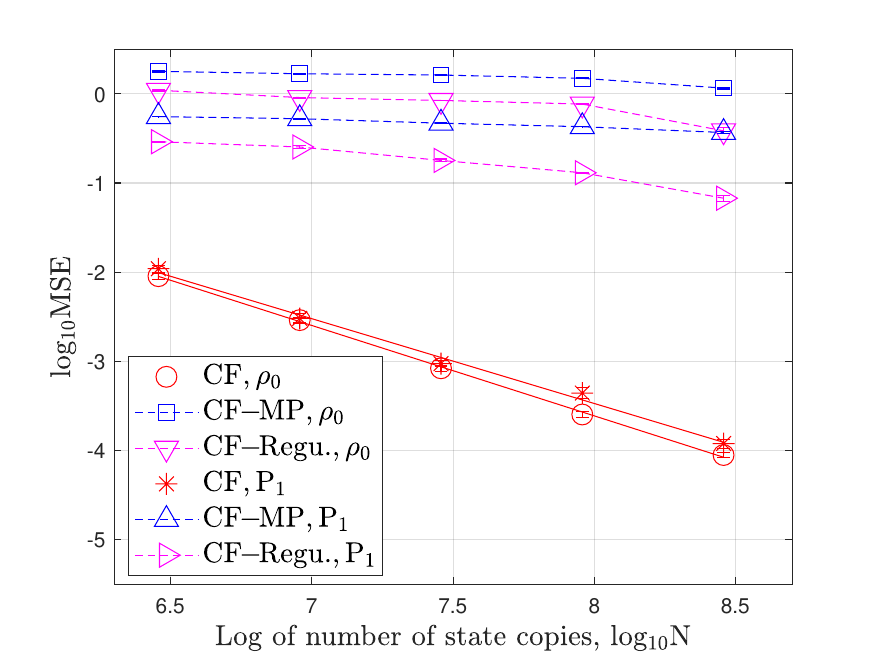}
		\end{minipage}
	}
	\subfigure{
		\begin{minipage}[b]{0.5\textwidth}
			\centering\includegraphics[width=1\textwidth]{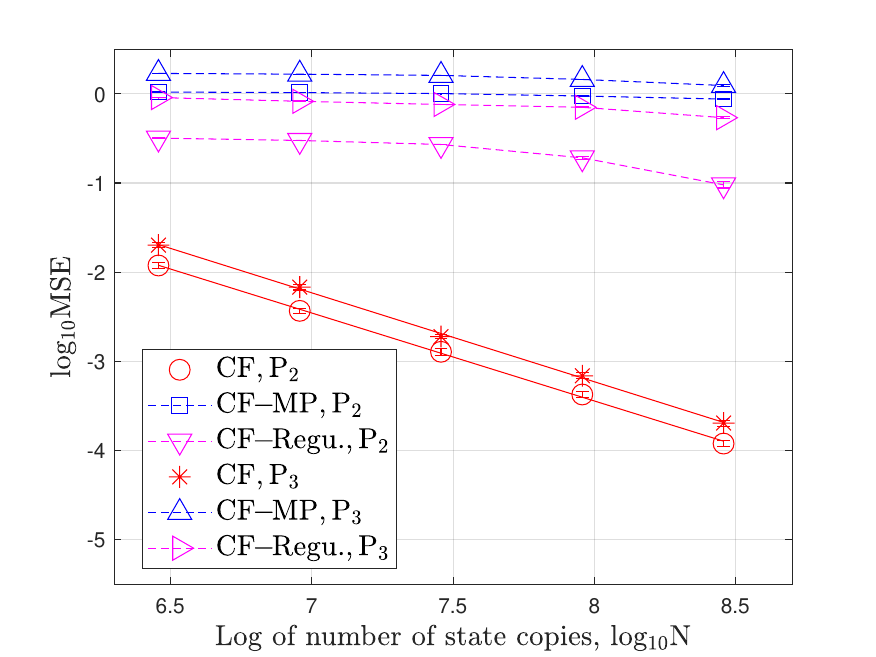}
		\end{minipage}
	}
	\caption{Log-log plot of MSE versus  the total number of copies $N$  for two-qubit mixed-unitary quantum processes using the closed-form (CF) solution in the informationally complete scenario, and using Moore–Penrose (MP) inverse  and  regularization (Regu.) in the informationally incomplete scenario.} \label{two}
\end{figure}

Due to the large number of optimization variables, SOS optimization proves to be time-consuming. Therefore, we opt to utilize only the closed-form solution in this scenario. We utilize all the $30$ mixed-unitary quantum processes which ensure informational completeness based on Problem \ref{problem1}. Additionally, we consider only the first $10$ pairs of Hamiltonians, resulting in an informationally incomplete scenario. The results are illustrated in Fig. \ref{two}, where both the QST and QDT exhibit MSE scalings of $O(1/N)$ in the informationally complete scenario. However, in the informationally incomplete scenario, the MSEs are considerably larger, and regularization outperforms MP inverse in terms of MSE.

\section{Conclusion}\label{sec7}
In this paper, we have proposed a framework to identify a quantum state and detector simultaneously using multiple quantum processes. We have designed a closed-form algorithm and proved that the MSE scalings of QST and QDT are both $O(1/N)$ in the informationally complete scenario.  We have also reformulated the problem as an SOS optimization problem. Moreover, we have discussed several illustrative examples including multiple Hamiltonians, mixed-unitary processes, and pure input states.
The numerical examples on one-qubit and two-qubit quantum systems have demonstrated the effectiveness of our close-formed solution and SOS optimization. Future work will focus on developing neural networks and shadow tomography in our framework.


\appendices

\section{Proof of  Proposition \ref{preal}}\label{ab}
\begin{IEEEproof}
	For each $q=[q_1, \cdots, q_{d^2}]^{T} \in \mathbb{R}^{d^2}$, let $Q\triangleq\sum_{i=1}^{d^2}q_i\Omega_i$.
For each $A\in \mathbb{C}^{d\times d}$, we have
	\begin{equation}\label{real1}
		\begin{aligned}
			U\operatorname{vec}(A^*\otimes A)U^\dagger q&=U\operatorname{vec}(A^*\otimes A)U^\dagger U\operatorname{vec}(Q)\\			
			&=U\operatorname{vec}(A^*\otimes A)\operatorname{vec}(Q)\\
			&= U\operatorname{vec}(AQA^\dagger).
		\end{aligned}
	\end{equation}
Since $AQA^{\dagger}$ is Hermitian, using \eqref{real}, we have $U\operatorname{vec}(AQA^\dagger)\in \mathbb{R}^{d^2}$, and thus $U\operatorname{vec}(A^*\otimes A)U^\dagger q \in \mathbb{R}^{d^2} $. Hence, let $q_k=[0,0, 1_{k}, 0\cdots,0]^{T}$ where only the $k$-th element is $1$, we have
\begin{equation}
	\begin{aligned}
	U\operatorname{vec}(A^*\otimes A)U^\dagger&= U\operatorname{vec}(A^*\otimes A)U^\dagger I\\
		 &= U\operatorname{vec}(A^*\otimes A)U^\dagger [q_1, \cdots, q_d] \in \mathbb{R}^{d^2 \times d^2}. 
		\end{aligned}
\end{equation}
\end{IEEEproof}

\section{Proof of  Theorem \ref{p3}}\label{ac}
\begin{IEEEproof}
	If the process $\mathcal{E}_a$ is generalized-unital and $\rho_0=I/{d}$,  we then have 
	\begin{equation}\label{unital}
		\left[\begin{array}{c}
			x_{a,0}\\
			0_{d^2-1}
		\end{array}\right]=	\left[\begin{array}{cc}
			r_a & t_a^{T} \\
			h_a & E_a
		\end{array}\right] \left[\begin{array}{c}
			1/\sqrt{d} \\
			0_{d^2-1}
		\end{array}\right],
	\end{equation}
	where $r_a=\sqrt{d}x_{a,0}=\operatorname{Tr}(\rho_a)$.
	Therefore, we have $\frac{1}{\sqrt{d}}h_a=0_{d^2-1}$ and thus $h_a=0_{d^2-1}$. Conversely, if $h_a=0_{d^2-1}$ and let $\rho_0=I/{d}$, using \eqref{unital}, we have $x_a=0_{d^2-1}$
	and thus $\rho_{a}=\alpha I/{d}$. Using Definition \ref{defuni}, the process is generalized-unital.
\end{IEEEproof}

\section{Proof of  Theorem \ref{th2}}\label{appc}
\begin{IEEEproof} 
	Let the parameterization matrix for the $j$-th group be
	\begin{equation}
		\begin{aligned}
			\mathcal{B}_a^{(j)}&\triangleq\sum_{i=1}^{d^2}\left(A_i^{a,(j)}\right)^* \otimes A_i^{a,(j)},\\
			\mathcal{B}^{(j)}&\triangleq\left[\operatorname{vec}\left(\mathcal{B}_1^{(j)}\right),  \cdots,\operatorname{vec}\left(\mathcal{B}_{L_j}^{(j)}\right)\right]^{T}.
		\end{aligned}
	\end{equation}
	We first consider the first group, i.e., $j=1$.
	Let $m=(u-1)d+l$ and $k=(v-1)d+h$ where $1\leq u,v,l,h \leq d$. The element at the $m$-th row and $k$-th column of $\mathcal{B}_a^{(1)}$ is 
	\begin{equation}\label{b1}
		\big(\mathcal{B}_a^{(1)}\big)_{mk}\!=\!\Big(\sum_{i=1}^{d^2}\left(A_i^{a,(1)}\right)^* \otimes A_i^{a,(1)}\Big)_{mk}\!\!=\!\sum_{i=1}^{d^2}\left(A_i^{a,(1)}\right)^{*}_{uv}  (A_i^{a,(1)})_{lh}.
	\end{equation}
	Denote $R_{j}\triangleq\mathcal{A}_{a}^{(j)}$. Thus $(R_1)_{vh}$ can be expressed as
	\begin{equation}\label{b2}
		(R_1)_{vh}\!=\!\Big(\sum_{i=1}^{d^2} ( A_{i}^{a,(1)})^{\dagger}  A_{i}^{a,(1)}\Big)_{vh}\!=\!\sum_{i=1}^{d^2} \Big(\sum_{u=1}^{d} (A_{i}^{a,(1)})^{*}_{uv} (A_{i}^{a,(1)})_{uh}\Big). 
	\end{equation}
	Therefore, using \eqref{b1} and \eqref{b2}, we have
	\begin{equation}
		\sum_{u=1}^{d} ({\mathcal{B}_a^{(1)}})_{(u-1)d+u, (v-1)d+h} =(R_1)_{vh}.
	\end{equation}
	Define matrices $\mathcal{B}^{R},\mathcal{B}_a^{(0)}\in \mathbb{C}^{d^2\times d^2}$ ($1\leq a\leq L_1$) as follows. The elements in $\mathcal{B}^{R}$ are all zero except $({\mathcal{B}^{R}})_{1, (v-1)d+h} =(R_1)_{vh}$ $\forall 1\leq v,h\leq d$. Moreover, $\mathcal{B}_a^{(0)}\triangleq \mathcal{B}_a^{(1)}-\mathcal{B}^{R}$. Hence, in each $\mathcal{B}_a^{(0)}$ ($1\leq a\leq L_{1}$) and for each $1\leq v,h\leq d$, there is a linear relationship
	\begin{equation}\label{eqW1}
		\sum_{u=1}^{d} ({\mathcal{B}_a^{(0)}})_{(u-1)d+u, (v-1)d+h} =0.
	\end{equation}
	We further define 
	\begin{equation}
		\mathcal{B}^{(0)}\triangleq\left[\operatorname{vec}\left(\mathcal{B}_1^{(0)}\right), \cdots,\operatorname{vec}\left(\mathcal{B}_{L_1}^{(0)}\right)\right]^{T},
	\end{equation}
	indicating that $({{\mathcal{B}}^{(1)}})^T=\left[\operatorname{vec}\left(\mathcal{B}_1^{(1)}\right), \cdots,\operatorname{vec}\left(\mathcal{B}_{L_1}^{(1)}\right)\right]=(\mathcal{B}^{(0)})^T+\left[\operatorname{vec}\left(\mathcal{B}^R\right), \cdots,\operatorname{vec}\left(\mathcal{B}^R\right)\right]$.
	
	Note that \eqref{eqW1} means there are $d^2$ linear-dependent constraints among the rows of $(\mathcal{B}^{(0)})^T$, and each row of $(\mathcal{B}^{(0)})^T$ appears at most once among these $d^2$ constraints. We thus know
	\begin{equation}
		\operatorname{rank}\big[(\mathcal{B}^{(0)})^T\big] \leq d^4-d^2.
	\end{equation}
	Therefore, we have
	\begin{equation}
		\begin{aligned}
			\operatorname{rank}(\mathcal{B}^{(1)})&=\operatorname{rank}(({\mathcal{B}^{(1)}})^T)\leq \operatorname{rank}\big[({\mathcal{B}^{(1)}})^T,\operatorname{vec}\left(\mathcal{B}^R\right)\big]\\
			&	=\operatorname{rank}\big[ (\mathcal{B}^{(0)})^T, \operatorname{vec}\left(\mathcal{B}^R\right)\big] \\
			&\leq d^4-d^2+1.
		\end{aligned}
	\end{equation}
	Similarly, we have $\operatorname{rank}(\mathcal{B}^{(j)}) \leq d^4-d^2+1$ for $1\leq j \leq f$.
	Since $\operatorname{rank}(\mathcal{B}^{(j)}) \leq L_j$,  for the $j$-th group of the processes, we have
	\begin{equation}
		\begin{aligned}
			\operatorname{rank}(\mathcal{B}^{(j)})&\leq \min \left(L_j, d^4-d^2+1\right).
		\end{aligned}
	\end{equation}
	Since $\mathcal{B}=[(\mathcal{B}^{(1)})^T, \cdots, (\mathcal{B}^{(f)})^T]^{T}$, we have
	\begin{equation}
		\begin{aligned}
			\operatorname{rank}(\mathcal{B})&\leq \sum_{j=1}^{f} \operatorname{rank}(\mathcal{B}^{(j)})\leq
			\sum_{j=1}^{f} \min \left(L_j, d^4-d^2+1\right).
		\end{aligned}
	\end{equation}
	Since $\mathcal{B}\in \mathbb{C}^{L\times d^4}$, we finally have
	\begin{equation}
		\begin{aligned}
			\operatorname{rank}(\mathcal{B})\leq  \min \Big(\sum_{j=1}^{f} \min \left(L_j, d^4-d^2+1\right), d^4\Big).
		\end{aligned}
	\end{equation}
\end{IEEEproof}

\section{Several lemmas}\label{appendixa}
\begin{lemma}(\cite{bhatia2007perturbation}, Theorem 8.1)\label{lemma1}
	Let $X$, $Y$ be Hermitian matrices with eigenvalues $\lambda_1(X)\geq\cdots\geq\lambda_n(X)$ and $\lambda_1(Y)\geq\cdots\geq\lambda_n(Y)$, respectively. Then
	\begin{equation}\label{weyl}
		\max_j|\lambda_j(X)-\lambda_j(Y)|\leq||X-Y||.
	\end{equation}
\end{lemma}

Let $\mathcal{X}, \mathcal{Y}$ be complex Euclidean spaces and $L(\mathcal{X}, \mathcal{Y})$ is referred to the collection of all linear mapping $A: \mathcal{X} \rightarrow \mathcal{Y}$. Define linear map $\Phi$ as $\Phi: L(\mathcal{X}) \rightarrow  L(\mathcal{Y})  $ and the set of all such maps is denoted as $T(\mathcal{X}, \mathcal{Y})$ \cite{watrous2018theory}. A map $\Phi$ is said to be Hermitian-preserving if it holds that $\Phi(H)$ is Hermitian for all $H$ is Hermitian \cite{watrous2018theory}.
\begin{lemma}(\cite{watrous2018theory}, Theorem 2.25) \label{lemma2}
	Let $\Phi \in T(\mathcal{X}, \mathcal{Y}) $ be a linear map for complex Euclidean space $\mathcal{X}$ and $\mathcal{Y}$. The following statements are equivalent:
	\begin{itemize}
		\item[(1)] $\Phi$ is Hermitian-preserving.
		\item[(2)] There exist completely positive maps $\Phi_0$, $\Phi_1$ for which $\Phi=\Phi_0-\Phi_1$.
	\end{itemize}
\end{lemma}

\section{Compressed sensing}\label{cs}
For compressed sensing in QST, Ref. \cite{Liu2011} proved that Pauli measurements satisfy the restricted isometry property (RIP) which have been implemented in \cite{Gross2010,Flammia2012}. Motivated by this, we also consider the Pauli unitary matrix  $V$ which is defined as  $V=\sigma_1\otimes\cdots\otimes\sigma_n$ where $\sigma_i\in\{I,\sigma_x,\sigma_y,\sigma_z\}$. Consider the following two Hermitian-preserving  processes
\begin{equation}\label{cs1}
	\begin{aligned}
		\mathcal{E}_1^{a}(\rho_0)&=\frac{g}{2}\left(V_1^{a} \rho_0 (V_2^{a})^{*} + (V_2^{a})^{*} \rho_0 V_1^{a}\right),\\
		\mathcal{E}_2^{a}(\rho_0)&=\frac{\mathrm{i}g}{2}\left(V_1^{a} \rho_0 (V_2^{a})^{*} - (V_2^{a})^{*} \rho_0 V_1^{a}\right),
	\end{aligned}
\end{equation}
where $V_1^{a}=(V_1^{a})^{\dagger}$ and $V_2^{a}=(V_2^{a})^{\dagger}$ are both Pauli unitary matrices, and $g\in \mathbb{R}$. Using Lemma \ref{lemma2} in Appendix \ref{appendixa}, we can find CP processes $(\mathcal{E}_1^{a})^{+}, (\mathcal{E}_1^{a})^{-}, (\mathcal{E}_2^{a})^{+}, (\mathcal{E}_2^{a})^{-}$ such that
\begin{equation}
	\begin{aligned}
		\mathcal{E}_1^{a}/g&=(\mathcal{E}_1^{a})^{+}-(\mathcal{E}_1^{a})^{-},\\
		\mathcal{E}_2/g&=(\mathcal{E}_2^{a})^{+}-(\mathcal{E}_2^{a})^{-}.
	\end{aligned}
\end{equation}
By choosing $g$ a positive number small enough, $g(\mathcal{E}_{i}^{a})^{j}$ can all be physically realizable for $i=1, 2$ and $j=+,-$.
If we input the same quantum state $\rho_0$ into these CP processes, let the measurement results of the $j$-th POVM element $P_j$ be
\begin{equation}
	\begin{aligned}
		p_{aj}^{1,+}=\operatorname{Tr}\Big(g(\mathcal{E}_1^{a})^{+}(\rho_{0})P_{j}\Big),  \;	p_{aj}^{1,-}=\operatorname{Tr}\Big(g(\mathcal{E}_1^{a})^{-}(\rho_{0})P_{j}\Big),\\
		p_{aj}^{2,+}=\operatorname{Tr}\Big(g(\mathcal{E}_2^{a})^{+}(\rho_{0})P_{j}\Big),  
		\;	p_{aj}^{2,-}=\operatorname{Tr}\Big(g(\mathcal{E}_2^{a})^{-}(\rho_{0})P_{j}\Big).
	\end{aligned}
\end{equation}
Define $\mathcal{E}_3^{a}\triangleq	\mathcal{E}_1^{a}-\mathrm{i}\mathcal{E}_2^{a} $. 
Therefore, we have
\begin{equation}
	\begin{aligned}
		\mathcal{E}_3^{a} (\rho_{0})=&\mathcal{E}_1^{a}(\rho_{0})-\mathrm{i}	\mathcal{E}_2^{a}(\rho_{0})=gV_1^{a} \rho_0 (V_2^{a})^{*} \\
		=&	g(\mathcal{E}_1^{a})^{+}(\rho_{0})-g(\mathcal{E}_1^{a})^{-}(\rho_{0})\\
		&-\mathrm{i}g(\mathcal{E}_2^{a})^{+}(\rho_{0})+\mathrm{i}g(\mathcal{E}_2^{a})^{-}(\rho_{0}).
	\end{aligned}
\end{equation}
Denote $$p_{aj}\triangleq\operatorname{Tr}\big(\mathcal{E}_3^{a}(\rho_{0})P_{j}\big)=p_{aj}^{1,+}-p_{aj}^{1,-}-\mathrm{i}p_{aj}^{2,+}+\mathrm{i}p_{aj}^{2,-}$$ which can also be expressed as
\begin{equation}
	\begin{aligned}
		p_{aj}&=g\operatorname{Tr}\Big(V_1^{a}\rho_{0}(V_2^{a})^{*} P_{j}\Big) \\
		&={g\left(\operatorname{vec}(V_{1}^{a}\otimes V_{2}^{a}) \right)^{\dagger}\left(\mathrm{vec}(P_{j})\otimes\operatorname{vec}(\rho_{0}^{T})\right)}.
	\end{aligned}
\end{equation}
Define $\mathcal{K}_j=\operatorname{vec}^{-1}(\mathrm{vec}(P_{j})\otimes\operatorname{vec}(\rho_{0}^{T}))$ and a linear map $\mathcal{T}_i: \mathbb{C}^{d^2 \times d^2}\rightarrow \mathbb{C}$ is introduced as:
\begin{equation}
	\begin{aligned}
		\hat p_{ij}&=g\operatorname{Tr}\big([V_{1}^{a}\otimes V_{2}^{a}] \mathcal{K}_j \big)+ e_{ij}\\
		&=\mathcal{T}_{i}(\mathcal{K}_j ) +e_{ij},
	\end{aligned}
\end{equation}
where $e_{ij}$ denotes statistical noise due to the finite number of samples. This equation closely resembles equation (2) in \cite{Flammia2012}.  
Since $V_{1}^{a}\otimes V_{2}^{a}$ are random Pauli unitary matrices which satisfy RIP \cite{Liu2011},
we can leverage the compressed sensing methodology outlined in \cite{Flammia2012} for our problem, particularly when both the quantum state and detector are of low rank, thereby reducing the sample complexity.

\section{Proof of  Proposition \ref{p1}}\label{ad}
\begin{IEEEproof}
	Refs. \cite{Merkel2010} and \cite{xiaoqst} have proved that the rank of the parameterization matrix generated by one Hamiltonian is not larger than $d^2-d+1$, i.e, $\operatorname{rank}(\mathcal{Q}_i)\leq d^2-d+1 $. Therefore,
	to ensure $B^{H}$ is full-rank, we need to prepare at least $\Big\lceil \frac{(d^2-1)^2}{d^2-d+1}\Big\rceil$ different Hamiltonians, i.e., $\mathcal{S}\geq \Big\lceil \frac{(d^2-1)^2}{d^2-d+1}\Big\rceil$.
\end{IEEEproof}

\bibliographystyle{ieeetr}         
\bibliography{qstdt} 

\begin{thebibliography}{10}

\bibitem{PhysRevLett.108.080502}
D.~Burgarth and K.~Yuasa, ``Quantum system identification,'' {\em Physical
  Review Letters}, vol.~108, p.~080502, 2012.

\bibitem{7130587}
M.~Gu{\c{t}}{\u{a}} and N.~Yamamoto, ``System identification for passive linear
  quantum systems,'' {\em IEEE Transactions on Automatic Control}, vol.~61,
  no.~4, pp.~921--936, 2016.

\bibitem{qci}
M.~A. Nielsen and I.~L. Chuang, {\em Quantum {C}omputation and {Q}uantum
  {I}nformation}.
\newblock Cambridge University Press, 2010.

\bibitem{dong2022quantum}
D.~Dong and I.~R. Petersen, ``Quantum estimation, control and learning:
  Opportunities and challenges,'' {\em Annual Reviews in Control}, vol.~54,
  pp.~243--251, 2022.

\bibitem{wiseman2009quantum}
H.~M. Wiseman and G.~J. Milburn, {\em Quantum {M}easurement and {C}ontrol}.
\newblock {C}ambridge {U}niversity {P}ress, 2009.

\bibitem{Dong2023}
D.~Dong and I.~R. Petersen, {\em Learning and {R}obust {C}ontrol in {Q}uantum
  {T}echnology}.
\newblock Springer Nature Switzerland AG, 2023.

\bibitem{RevModPhys.89.035002}
C.~L. Degen, F.~Reinhard, and P.~Cappellaro, ``Quantum sensing,'' {\em Reviews
  of Modern Physics}, vol.~89, p.~035002, 2017.

\bibitem{Zhang2017}
J.~Zhang, Y.-X. Liu, R.-B. Wu, K.~Jacobs, and F.~Nori, ``Quantum feedback:
  Theory, experiments, and applications,'' {\em Physics Reports}, vol.~679,
  pp.~1--60, 2017.

\bibitem{Liang2024}
W.~Liang, T.~Grigoletto, and F.~Ticozzi, ``Dissipative feedback switching for
  quantum stabilization,'' {\em Automatica}, vol.~165, p.~111659, 2024.

\bibitem{qstmle}
Z.~Hradil, ``Quantum-state estimation,'' {\em Physical Review A}, vol.~55,
  pp.~R1561--R1564, 1997.

\bibitem{effqst}
J.~A. Smolin, J.~M. Gambetta, and G.~Smith, ``Efficient method for computing
  the maximum-likelihood quantum state from measurements with additive
  {G}aussian noise,'' {\em Physical Review Letters}, vol.~108, p.~070502, 2012.

\bibitem{Qi2013}
B.~Qi, Z.~Hou, L.~Li, D.~Dong, G.-Y. Xiang, and G.-C. Guo, ``Quantum state
  tomography via linear regression estimation,'' {\em Scientific Reports},
  vol.~3, p.~3496, 2013.

\bibitem{xiaorank}
S.~Xiao, Y.~Wang, J.~Zhang, D.~Dong, and H.~Yonezawa, ``Quantum state and
  detector tomography with known rank,'' {\em IFAC-PapersOnLine}, vol.~56,
  no.~2, pp.~5881--5887, 2023.

\bibitem{Gross2010}
D.~Gross, Y.-K. Liu, S.~T. Flammia, S.~Becker, and J.~Eisert, ``Quantum state
  tomography via compressed sensing,'' {\em Physical Review Letters}, vol.~105,
  p.~150401, 2010.

\bibitem{Flammia2012}
S.~T. Flammia, D.~Gross, Y.-K. Liu, and J.~Eisert, ``Quantum tomography via
  compressed sensing: error bounds, sample complexity and efficient
  estimators,'' {\em New Journal of Physics}, vol.~14, no.~9, p.~095022, 2012.

\bibitem{MU2020108837}
B.~Mu, H.~Qi, I.~R. Petersen, and G.~Shi, ``Quantum tomography by regularized
  linear regressions,'' {\em Automatica}, vol.~114, p.~108837, 2020.

\bibitem{Aaronson2018}
S.~Aaronson, ``Shadow tomography of quantum states,'' in {\em Proceedings of
  the 50th Annual ACM SIGACT Symposium on Theory of Computing}, STOC 2018,
  p.~325–338, 2018.

\bibitem{Huang2020}
H.-Y. Huang, R.~Kueng, and J.~Preskill, ``Predicting many properties of a
  quantum system from very few measurements,'' {\em Nature Physics}, vol.~16,
  no.~10, pp.~1050--1057, 2020.

\bibitem{PhysRevA.64.024102}
J.~Fiur\'a\ifmmode~\check{s}\else \v{s}\fi{}ek, ``Maximum-likelihood estimation
  of quantum measurement,'' {\em Physical Review A}, vol.~64, p.~024102, 2001.

\bibitem{Grandi_2017}
S.~Grandi, A.~Zavatta, M.~Bellini, and M.~G.~A. Paris, ``Experimental quantum
  tomography of a homodyne detector,'' {\em New Journal of Physics}, vol.~19,
  no.~5, p.~053015, 2017.

\bibitem{Renema2012}
J.~J. Renema, G.~Frucci, Z.~Zhou, F.~Mattioli, A.~Gaggero, R.~Leoni, M.~J.~A.
  de~Dood, A.~Fiore, and M.~P. van Exter, ``Modified detector tomography
  technique applied to a superconducting multiphoton nanodetector,'' {\em
  Optics {E}xpress}, vol.~20, no.~3, pp.~2806--2813, 2012.

\bibitem{Feito_2009}
A.~Feito, J.~S. Lundeen, H.~Coldenstrodt-Ronge, J.~Eisert, M.~B. Plenio, and
  I.~A. Walmsley, ``Measuring measurement: theory and practice,'' {\em New
  Journal of Physics}, vol.~11, no.~9, p.~093038, 2009.

\bibitem{Lundeen2009}
J.~S. Lundeen, A.~Feito, H.~Coldenstrodt-Ronge, K.~L. Pregnell, C.~Silberhorn,
  T.~C. Ralph, J.~Eisert, M.~B. Plenio, and I.~A. Walmsley, ``Tomography of
  quantum detectors,'' {\em Nature Physics}, vol.~5, no.~1, pp.~27--30, 2009.

\bibitem{Zhang_2012}
L.~Zhang, A.~Datta, H.~B. Coldenstrodt-Ronge, X.-M. Jin, J.~Eisert, M.~B.
  Plenio, and I.~A. Walmsley, ``Recursive quantum detector tomography,'' {\em
  New Journal of Physics}, vol.~14, no.~11, p.~115005, 2012.

\bibitem{wang2019twostage}
Y.~{Wang}, S.~{Yokoyama}, D.~{Dong}, I.~R. {Petersen}, E.~H. {Huntington}, and
  H.~{Yonezawa}, ``Two-stage estimation for quantum detector tomography: Error
  analysis, numerical and experimental results,'' {\em IEEE Transactions on
  Information Theory}, vol.~67, no.~4, pp.~2293--2307, 2021.

\bibitem{xiao2021optimal}
S.~Xiao, Y.~Wang, D.~Dong, and J.~Zhang, ``Optimal and two-step adaptive
  quantum detector tomography,'' {\em Automatica}, vol.~141, p.~110296, 2022.

\bibitem{Xiao2023}
S.~Xiao, Y.~Wang, J.~Zhang, D.~Dong, S.~Yokoyama, I.~R. Petersen, and
  H.~Yonezawa, ``On the regularization and optimization in quantum detector
  tomography,'' {\em Automatica}, vol.~155, p.~111124, 2023.

\bibitem{PhysRevLett.124.040402}
A.~Zhang, J.~Xie, H.~Xu, K.~Zheng, H.~Zhang, Y.-T. Poon, V.~Vedral, and
  L.~Zhang, ``Experimental self-characterization of quantum measurements,''
  {\em Physical Review Letters}, vol.~124, p.~040402, 2020.

\bibitem{PhysRevLett.127.180401}
L.~Xu, H.~Xu, T.~Jiang, F.~Xu, K.~Zheng, B.~Wang, A.~Zhang, and L.~Zhang,
  ``Direct characterization of quantum measurements using weak values,'' {\em
  Physical Review Letters}, vol.~127, p.~180401, 2021.

\bibitem{PhysRevA.98.042318}
A.~C. Keith, C.~H. Baldwin, S.~Glancy, and E.~Knill, ``Joint quantum-state and
  measurement tomography with incomplete measurements,'' {\em Physical Review
  A}, vol.~98, p.~042318, 2018.

\bibitem{PhysRevA.104.012416}
A.~Stephens, J.~M. Cutshall, T.~McPhee, and M.~Beck, ``Self-consistent state
  and measurement tomography with fewer measurements,'' {\em Physical Review
  A}, vol.~104, p.~012416, 2021.

\bibitem{sostools}
A.~Papachristodoulou, J.~Anderson, G.~Valmorbida, S.~Prajna, P.~Seiler,
  P.~Parrilo, M.~M. Peet, and D.~Jagt, ``{SOSTOOLS} version 4.00 sum of squares
  optimization toolbox for {MATLAB},'' {\em arXiv:1310.4716}, 2021.

\bibitem{watrous2018theory}
J.~Watrous, {\em The {T}heory of {Q}uantum {I}nformation}.
\newblock {C}ambridge {U}niversity {P}ress, 2018.

\bibitem{sic}
J.~M. Renes, R.~Blume-Kohout, A.~J. Scott, and C.~M. Caves, ``Symmetric
  informationally complete quantum measurements,'' {\em Journal of Mathematical
  Physics}, vol.~45, no.~6, pp.~2171--2180, 2004.

\bibitem{mubreview}
T.~Durt, B.-G. Englert, I.~Bengtsson, and K.~Życzkowski, ``On mutually
  unbiased bases,'' {\em International Journal of Quantum Information},
  vol.~08, no.~04, pp.~535--640, 2010.

\bibitem{PhysRevA.78.052122}
M.~D. de~Burgh, N.~K. Langford, A.~C. Doherty, and A.~Gilchrist, ``Choice of
  measurement sets in qubit tomography,'' {\em Physical Review A}, vol.~78,
  p.~052122, 2008.

\bibitem{Mendl2009}
C.~B. Mendl and M.~M. Wolf, ``Unital quantum channels--convex structure and
  revivals of {B}irkhoff’s theorem,'' {\em Communications in Mathematical
  Physics}, vol.~289, no.~3, pp.~1057--1086, 2009.

\bibitem{zhang1}
J.~Zhang and M.~Sarovar, ``Quantum {H}amiltonian identification from
  measurement time traces,'' {\em Physical Review Letters}, vol.~113, no.~8,
  p.~080401, 2014.

\bibitem{alicki}
R.~Alicki and K.~Lendi, {\em Quantum {D}ynamical {S}emigroups and
  {A}pplications}, vol.~717.
\newblock New York: Springer, 2007.

\bibitem{8022944}
Y.~{Wang}, D.~{Dong}, B.~{Qi}, J.~{Zhang}, I.~R. {Petersen}, and H.~{Yonezawa},
  ``A quantum {H}amiltonian identification algorithm: Computational complexity
  and error analysis,'' {\em IEEE Transactions on Automatic Control}, vol.~63,
  no.~5, pp.~1388--1403, 2018.

\bibitem{zhang2}
J.~Zhang and M.~Sarovar, ``Identification of open quantum systems from
  observable time traces,'' {\em Physical Review A}, vol.~91, p.~052121, 2015.

\bibitem{Gebhart2023}
V.~Gebhart, R.~Santagati, A.~A. Gentile, E.~M. Gauger, D.~Craig, N.~Ares,
  L.~Banchi, F.~Marquardt, L.~Pezzè, and C.~Bonato, ``Learning quantum
  systems,'' {\em Nature Reviews Physics}, vol.~5, no.~3, pp.~141--156, 2023.

\bibitem{PhysRevA.77.012307}
E.~Knill, D.~Leibfried, R.~Reichle, J.~Britton, R.~B. Blakestad, J.~D. Jost,
  C.~Langer, R.~Ozeri, S.~Seidelin, and D.~J. Wineland, ``Randomized
  benchmarking of quantum gates,'' {\em Physical Review A}, vol.~77, p.~012307,
  2008.

\bibitem{PRXQuantum.3.020357}
J.~Helsen, I.~Roth, E.~Onorati, A.~Werner, and J.~Eisert, ``General framework
  for randomized benchmarking,'' {\em PRX Quantum}, vol.~3, p.~020357, 2022.

\bibitem{6883125}
T.~Chen, M.~S. Andersen, L.~Ljung, A.~Chiuso, and G.~Pillonetto, ``System
  identification via sparse multiple kernel-based regularization using
  sequential convex optimization techniques,'' {\em IEEE Transactions on
  Automatic Control}, vol.~59, no.~11, pp.~2933--2945, 2014.

\bibitem{Chen2018}
T.~Chen, ``On kernel design for regularized {LTI} system identification,'' {\em
  Automatica}, vol.~90, pp.~109--122, 2018.

\bibitem{10273596}
B.~Mu and T.~Chen, ``On asymptotic optimality of cross-validation estimators
  for kernel-based regularized system identification,'' {\em IEEE Transactions
  on Automatic Control}, vol.~69, no.~7, pp.~4352--4367, 2024.

\bibitem{LOAN200085}
C.~F. Loan, ``The ubiquitous {K}ronecker product,'' {\em Journal of
  Computational and Applied Mathematics}, vol.~123, no.~1, pp.~85--100, 2000.

\bibitem{8263706}
A.~A. Ahmadi, G.~Hall, A.~Papachristodoulou, J.~Saunderson, and Y.~Zheng,
  ``Improving efficiency and scalability of sum of squares optimization: Recent
  advances and limitations,'' in {\em 2017 IEEE 56th Annual Conference on
  Decision and Control (CDC)}, pp.~453--462, 2017.

\bibitem{Boyd2004Convex}
S.~Boyd and L.~Vandenberghe, {\em Convex Optimization}.
\newblock Cambridge University Press, 2004.

\bibitem{parrilo2001}
P.~A. Parrilo and B.~Sturmfels, ``Minimizing polynomial functions,'' {\em In
  Algorithmic and quantitative real algebraic geometry, DIMACS Series in
  Discrete Mathematics and Theoretical Computer Science, AMS}, vol.~60,
  pp.~83--99, 2003.

\bibitem{Kimura2003}
G.~Kimura, ``The {B}loch vector for {N}-level systems,'' {\em Physics Letters
  A}, vol.~314, no.~5, pp.~339--349, 2003.

\bibitem{1440563}
R.~van Handel, J.~Stockton, and H.~Mabuchi, ``Feedback control of quantum state
  reduction,'' {\em IEEE Transactions on Automatic Control}, vol.~50, no.~6,
  pp.~768--780, 2005.

\bibitem{Merkel2010}
S.~T. Merkel, C.~A. Riofr\'{\i}o, S.~T. Flammia, and I.~H. Deutsch, ``Random
  unitary maps for quantum state reconstruction,'' {\em Physical Review A},
  vol.~81, p.~032126, 2010.

\bibitem{xiaoqst}
S.~Xiao, Y.~Wang, Q.~Yu, J.~Zhang, D.~Dong, and I.~R. Petersen, ``Quantum state
  tomography from observable time traces in closed quantum systems,'' {\em
  Control Theory and Technology}, vol.~22, no.~2, pp.~222--234, 2024.

\bibitem{Cole2005}
J.~H. Cole, S.~G. Schirmer, A.~D. Greentree, C.~J. Wellard, D.~K.~L. Oi, and
  L.~C.~L. Hollenberg, ``Identifying an experimental two-state {H}amiltonian to
  arbitrary accuracy,'' {\em Physical Review A}, vol.~71, p.~062312, 2005.

\bibitem{Girard2022}
M.~Girard, D.~Leung, J.~Levick, C.-K. Li, V.~Paulsen, Y.~T. Poon, and
  J.~Watrous, ``On the mixed-unitary rank of quantum channels,'' {\em
  Communications in Mathematical Physics}, vol.~394, no.~2, pp.~919--951, 2022.

\bibitem{MISZCZAK2012118}
J.~A. Miszczak, ``Generating and using truly random quantum states in
  {M}athematica,'' {\em Computer Physics Communications}, vol.~183, no.~1,
  pp.~118--124, 2012.

\bibitem{qetlab}
N.~Johnston, ``{QETLAB}: A {MATLAB} toolbox for quantum entanglement, version
  0.9,'' Jan. 2016.

\bibitem{bhatia2007perturbation}
R.~Bhatia, {\em {P}erturbation {B}ounds for {M}atrix {E}igenvalues}.
\newblock {S}ociety for {I}ndustrial and {A}pplied {M}athematics, 2007.

\bibitem{Liu2011}
Y.-K. Liu, ``Universal low-rank matrix recovery from {P}auli measurements,'' in
  {\em Advances in Neural Information Processing Systems}, vol.~24, 2011.

\end{thebibliography}

\end{document}